\documentclass[a4paper,english,11pt]{article}
\usepackage{epsf,amsmath,amssymb,graphicx,dcolumn}
\usepackage{scalefnt,subfigure,ulem,pstricks}
\usepackage{cite,hyperref}
\usepackage{todonotes}
\usepackage{color}
\usepackage{rotating}
\renewcommand{\arraystretch}{1.2}
\newcommand{\fs}[1]{{\abbrev #1FS}}
\newcommand{\pdf}{{\abbrev PDF}}
\newcommand{\qcd}{{\abbrev QCD}}

\newcommand{\abbrev}{\scalefont{.9}}

\newcommand{\muF}{\mu_\text{F}}
\newcommand{\muR}{\mu_\text{R}}
\newcommand{\mhiggs}{m_H}

\newcommand{\mbottom}{m_b}
\newcommand{\ep}{\epsilon}

\newcommand{\eqn}[1]{Eq.\,(\ref{#1})}
\newcommand{\fig}[1]{Fig.\,\ref{#1}}

\newcommand{\order}[1]{{\cal O}(#1)}
\newcommand{\lhc}{{\abbrev LHC}}
\newcommand{\sm}{{\abbrev SM}}
\newcommand{\mssm}{{\abbrev MSSM}}
\newcommand{\susy}{{\abbrev SUSY}}

\newcommand{\lo}{\text{\abbrev LO}}
\newcommand{\nlo}{\text{\abbrev NLO}}
\newcommand{\nnlo}{\text{\abbrev NNLO}}

\newcommand{\msbar}{\overline{\mbox{\abbrev MS}}}

\newcommand{\bld}[1]{\boldmath{$#1$}}

\title{\vspace*{-6em}
  \begin{flushright}
    {\sf\small November 2011 --- CERN-PH-TH/2011-269 --- WUB/11-19}
  \end{flushright}
  \vspace*{2em} Jet-veto in bottom-quark induced Higgs production at
  next-to-next-to-leading order}
\author{Robert V. Harlander and Marius Wiesemann\\[2em] {\it Fachbereich C,
    Bergische Universit\"at Wuppertal}\\{\it 42097 Wuppertal,
    Germany}\\ {\small\tt harlander@physik.uni-wuppertal.de}\\[-.3em]
  {\small\tt m.wiesemann@uni-wuppertal.de} }
\date{}
\begin{document}
\maketitle


\begin{abstract}
We present results for associated Higgs+$n$-jet production in bottom
quark annihilation, for $n=0$ and $n\geq 1$ at \nnlo{} and \nlo{}
accuracy, respectively. We consider both the cases with and without
$b$-tagging. Numerical results are presented for parameters relevant for
experiments at the \lhc{}.
\end{abstract}


\section{Introduction}\label{sec::intro}

The Standard Model (\sm{}) and its supersymmetric extensions require a
mechanism to explain the gauge boson and fermion masses. The Higgs
mechanism where the particles acquire masses through interactions with
the Higgs field(s) is the most popular ansatz. One of the primary goals
of the Large Hadron Collider (\lhc{}) is to find or exclude a Higgs
boson over the full theoretically meaningful mass range.

Various channels can be exploited for the search of Higgs bosons at
hadron colliders (see, e.g.,
Refs.~\cite{Djouadi:2005gi,Djouadi:2005gj}). While in the \sm{} gluon
fusion has the largest cross section by far, in supersymmetric (\susy{})
theories the radiation off bottom-quarks becomes
equally important (see, e.g., Ref.~\cite{Belyaev:2005ct}):
\begin{align}
pp \rightarrow (b\bar{b})H+X\,.
\label{eq::ppbbh}
\end{align}
Here and in what follows, $H$ denotes a generic neutral Higgs boson,
scalar or pseudo-scalar. In particular, it includes the light and heavy
{\abbrev CP}-even as well as the {\abbrev CP}-odd neutral Higgs bosons
of the Minimal Supersymmetric Standard Model (\mssm{}), $h^0,H^0$, and
$A$.

\begin{figure}[tbp]
\centering
\subfigure[]{\label{fig::four}\includegraphics[width=6em]{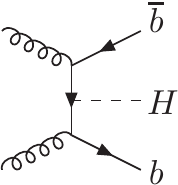}}
\hspace{1.5in}
\subfigure[]{\label{fig::five}\includegraphics[width=7.5em]{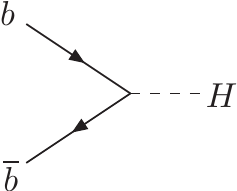}}
\caption{Leading order diagrams for the associated production of a Higgs
  with bottom-quarks in the (a)~four- and (b)~five-flavor scheme.}
\end{figure}

Two approaches have been pursued in the literature to calculate the
process (\ref{eq::ppbbh}). In the so-called ``four-flavor scheme''
(\fs{4}), the leading order (\lo{}) processes are
\begin{align}
gg/q\bar{q}\rightarrow b\bar{b}H\,,
\end{align}
where $q\in\{u,d,s,c\}$. The collinear region of the bottom-quark
momenta occurring in the $gg$ initiated process, see \fig{fig::four}, is
regulated by the bottom-quark mass $\mbottom$ and leads to potentially
large logarithms $\ln\left(\muF^2/\mbottom^2\right)$, where $\muF\simeq
\mhiggs$ denotes the factorization scale, and $\mhiggs$ the Higgs mass. In
this approach, the total cross section for a scalar Higgs boson is known
to next-to-leading order (\nlo) \qcd{}
accuracy~\cite{Dittmaier:2003ej,Dawson:2003kb}; \susy{} effects have
been considered as well~\cite{Gao:2004wg}.


The other approach to calculate the cross section for the process in
\eqn{eq::ppbbh} is the so-called ``five-flavor scheme'' (\fs{5}), where
the \lo{} partonic reaction is
\begin{align}
b\bar{b}\rightarrow H\,.
\end{align}
The corresponding Feynman diagram is shown in Fig.\,\ref{fig::five}.
The {\abbrev DGLAP} evolution of the bottom-quark parton densities
formally resums the collinear logarithms that are manifest in the
\fs{4}, see above, leading to a better perturbative convergence.
However, effects from bottom-quark production at large transverse
momentum $p_T$ are taken into account only at higher orders in the
\fs{5}. In fact, the next-to-next-to-leading order (\nnlo{})
prediction~\cite{Harlander:2003ai} plays a special role in this process
because only from this order on, the \fs{5} approach includes the \lo{}
diagram of the \fs{4} (see Ref.~\cite{Harlander:2003ai} for a more
detailed discussion). In the \fs{5}, also electro-weak corrections have
been evaluated~\cite{Dittmaier:2006cz}.

Both the \fs{4} and \fs{5} are formally viable approaches to calculate
the inclusive cross section for the process shown in
\eqn{eq::ppbbh}. Nevertheless, it took a significant amount of efforts
to pin down their qualitative and quantitative differences (see, e.g.,
Refs.\cite{Rainwater:2002hm,Plehn:2002vy,Dicus:1998hs,Harlander:2003ai}).
The ``\lhc{} Higgs Cross Section Working Group'' \cite{Dittmaier:2011ti}
has now decided to combine the inclusive cross sections of both schemes
according to the so-called ``Santander Matching''
procedure~\cite{bbh-santander}. In order to optimally exploit the
advantages of each approach in its region of applicability, they enter
the cross section prediction with a Higgs-mass dependent weight in this
procedure.

In summary, the inclusive Higgs cross section in bottom-quark
annihilation is under good theoretical control. However, it is well
known that exclusive $H+$jet production can be advantageous for
experimental analyses. In gluon fusion, this process has been studied in
quite some detail, both in the
\sm{}~\cite{deFlorian:1999zd,Glosser:2002gm,Ravindran:2002dc,deFlorian:2000pr,
  Catani:2001cr,Berger:2002ut,Bozzi:2005wk,Anastasiou:2004xq,Anastasiou:2005qj,
  Campbell:2006xx,Catani:2007vq,Bozzi:2007pn,Berger:2010xi} and the
Minimal Supersymmetric Standard Model
(\mssm{})~\cite{Field:2002pb,Field:2003yy,Langenegger:2006wu,Brein:2003df,
  Brein:2007da,Brein:2010xj}.  As mentioned above, in the \mssm{} it is
essential to take $(b\bar{b})H+$jet production into account as well. In
this paper we present the corresponding cross sections in the framework
of the \fs{5}. In particular, we show results for the Higgs plus 0- and
$\geq 1$-jet cross sections for Higgs production in bottom-quark
annihilation at \nnlo{} and \nlo{}, respectively.  The results are given
in the \sm{}, but according to the studies of
Refs.~\cite{Dawson:2011pe,Dittmaier:2006cz}, they are applicable to the
\mssm{} by simply rescaling the $b\bar{b}H$ coupling.

Let us stress at this point the difference between this study and
similar ones existing in the literature. In
Ref.~\cite{Harlander:2010cz}, we calculated kinematical distributions of
the Higgs boson in $H+$jet production, while here we focus on the aspect
specific to the associated jets. Ref.~\cite{Campbell:2002zm} considered
the \nlo{} cross section for Higgs production in association with one or
two {\it bottom} jets ($H+nb$-jets, $n=1,2$), which is contained in our
calculation and which we used as an important check. The analogous study
in the \fs{4} has been performed in Ref.~\cite{Dawson:2004sh}.  Our
viewpoint here is simply that bottom-induced Higgs production may
actually be the dominant mechanism for Higgs production, so that the
studies done for $H+$jet production in gluon fusion should be
supplemented by the bottom-annihilation contribution. Nevertheless, we
will also include updated numbers for $H+nb$-jet production ($n=1,2$) in
this paper, and present \nnlo{} results for a $b$-jet veto, i.e.\ for
$H+0b$-jet production. One caveat should be added, however, namely that
for a fully consistent combination of gluon fusion and $b\bar bH$
production, one would have to include interference terms of these two
processes which will be neglected in what follows (as it has been done
in all existing studies of the inclusive cross section up to now).

The remainder of this paper is organized as follows: In
Section~\ref{sec::calc}, we describe our calculation of the $H+n$-jet
($n=0$ and $n\geq 1$) and the $H+nb$-jet cross section
($n=0,1,2$). Section~\ref{sec::results} contains our results for a
default set of parameters. In particular, we consider proton-proton
collisions at 7\,TeV, while the numbers for 14\,TeV center-of-mass
energy are collected to Appendix~\ref{app::14}. Our conclusions are
presented in Section~\ref{sec::concl}.


\section{Calculation of \bld{(b\bar{b})H+}jet production}
\label{sec::calc}



\begin{figure}
\centering
\subfigure[]{\label{fig::lead}\includegraphics[width=11.5em]{%
    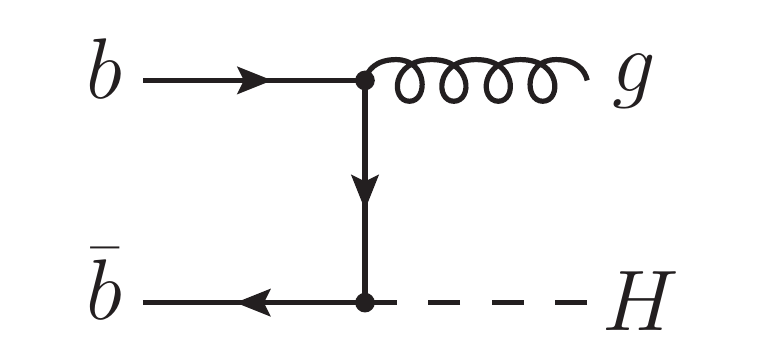}}
\hspace{1.5in}
\subfigure[]{\label{fig::leadgb}\includegraphics[width=11.5em]{%
    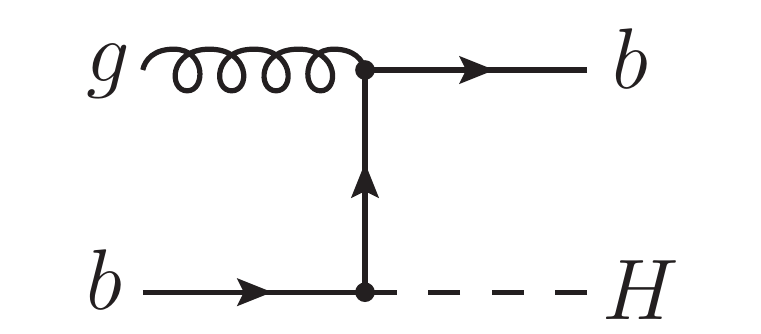}}
\caption[]{\label{fig::LO}
  Representative diagrams of the leading order channels
  (a)~$b\bar{b}\rightarrow gH$ and (b)~$bg \rightarrow bH$.}
\end{figure}


\begin{figure}[tbp]
\centering
\subfigure[]{\label{fig::bbloop}\includegraphics[width=11.5em]{%
    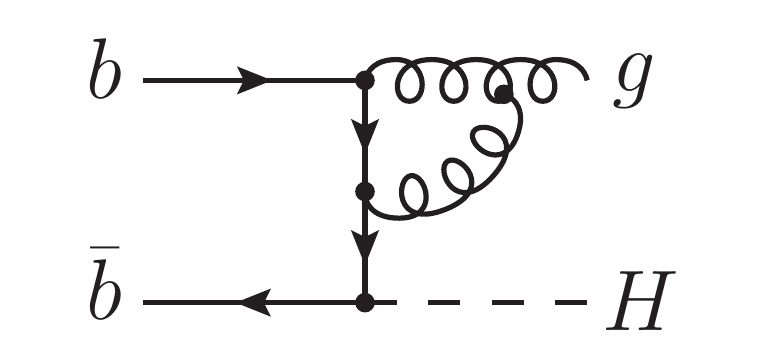}}
\hspace{1.5in}
\subfigure[]{\label{fig::bgloop}\includegraphics[width=11.5em]{%
    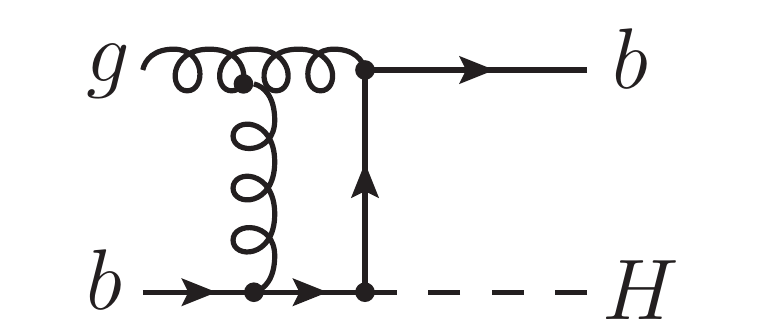}}\\
\subfigure[]{\label{fig::bbggH}\includegraphics[width=11.5em]{%
    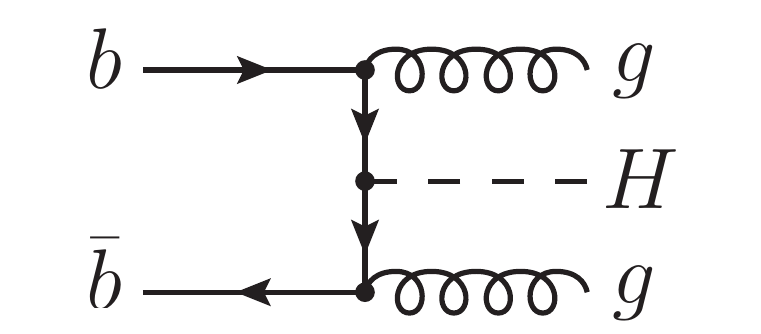}}
\hspace{1.5in}
\subfigure[]{\label{fig::bqbqH}\includegraphics[width=11.5em]{%
    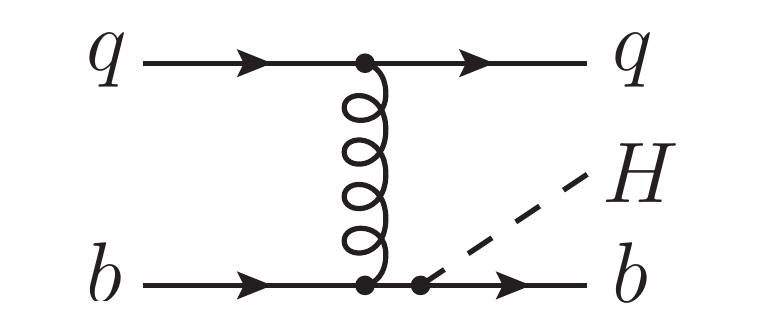}}
\caption[]{\label{fig::NLO}Representative diagrams at \nlo{}: (a),(b)
  virtual and (c),(d) real corrections.}
\end{figure}


Considering $H+$jet production in bottom-quark annihilation at \nlo{},
several subprocesses have to be taken into account. The generic leading
order channels are $b\bar{b}\rightarrow gH$ and $gb\rightarrow bH$, see
\fig{fig::LO}.  In \fig{fig::bbloop} and \ref{fig::bgloop}, two
representative diagrams of the corresponding virtual corrections to
$b\bar{b}\rightarrow gH$ and $gb\rightarrow bH$ are shown.  The real
emission processes derived from the \lo{} channels are
$b\bar{b}\rightarrow ggH$, $b\bar{b}\rightarrow b\bar{b}H$,
$b\bar{b}\rightarrow q\bar{q}H$, and $gb\rightarrow bgH$.  In addition,
the sub-channels $gg\rightarrow b\bar{b}H$, $qb\rightarrow qbH$,
$bb\rightarrow bbH$, and $q\bar{q}\rightarrow b\bar{b}$ ($q\in \{u,d,
s,c\}$) which do not have a \lo{} correspondence need to be taken into
account.  Two representative real emission diagrams are displayed in
Fig.\,\ref{fig::bbggH} and \ref{fig::bqbqH}. It is understood that the
charge conjugated processes must be included as well.  We note that
diagrams where the Higgs boson is radiated off of closed top or bottom
quark loops are usually attributed to the gluon fusion channel, and
therefore are not taken into account here. As already pointed out in the
introduction, for a consistent treatment of the $H+n$-jet cross section
these two processes should be combined, and also interference terms
should be taken into account.

The calculation was carried out using the program described in
Ref.~\cite{Harlander:2010cz}, where we implemented the anti-$k_T$
jet-algorithm\footnote{Since at most two jets can occur at the order we
  are considering, the anti-$k_T$ leads to the same results as the $k_T$
  and the Cambridge-Aachen algorithm.}~\cite{Cacciari:2008gp} to
identify \qcd{} jets.


\subsection{Cross section without jet-flavor tagging}\label{sec:calc:notag}

Our setup allows us to calculate the exclusive and inclusive $H+n$-jet
cross sections $\sigma_{n\text{-jet}}$ and $\sigma_{\ge n\text{-jet}}$,
respectively, where for $n=1$ we work at \nlo{} accuracy, while for
$n=2$ we only get a \lo{} result.  (The inclusive and exclusive
$H+2$-jet cross section are identical at this order, $\sigma_{\geq
  \text{2-jet}} = \sigma_\text{2-jet} + \order{\alpha_s^3}$, where
$\alpha_s$ is the strong coupling constant.) Since our
results do not include any parton showering or hadronization, ``jet''
denotes any outgoing quark, anti-quark or gluon, irrespective of the
quark flavor. Further below, we will also consider the case of
$b$-tagged cross sections.

With the knowledge of the total inclusive cross section
$\sigma_{\text{tot}}$ up to \nnlo{}~\cite{Harlander:2003ai} we can use
our program for the inclusive $H+1$-jet rate (also referred to as
$H+\geq1$-jet or inclusive $H+$jet rate in the following) at \nlo{} to
obtain the exclusive $H+0$-jet, or jet-vetoed, cross section at \nnlo{}:
\begin{align}
\sigma_\text{jet-veto}^\nnlo{} \equiv
\sigma^\nnlo_{0\text{-jet}}=\sigma^\nnlo_{\text{tot}}
-\sigma^{\nlo'}_{\ge 1\text{-jet}}\,.
\label{eq::0jet}
\end{align}
Since this quantity is formally of \nnlo{}, both contributions on the
right side of \eqn{eq::0jet} have to be calculated with \nnlo{} parton
density functions (\pdf{}s) and couplings. This is indicated by the
prime in $\sigma^{\nlo'}_{\ge 1\text{-jet}}$ which distinguishes it from
the proper \nlo{} quantity $\sigma^\nlo_{\ge 1\text{-jet}}$. Similar to
the fact that $\sigma^\nlo_{1\text{-jet}} + \sigma^\lo_{2\text{-jet}}
\not\equiv \sigma^\nlo_{\ge 1\text{-jet}} (\equiv
\sigma^\nlo_{1\text{-jet}} +\sigma^{\lo'}_{2\text{-jet}})$ due to
the different orders of \pdf{}s and couplings that need to be used, it
is clear that
\begin{equation}
\begin{split}
\sigma^\nnlo_{0\text{-jet}}
+\sigma^\nlo_{\ge1\text{-jet}}
\not\equiv
\sigma^\nnlo_{\text{tot}} \not\equiv
\sigma^\nnlo_{0\text{-jet}}
+\sigma^\nlo_{1\text{-jet}}
+\sigma^\lo_{2\text{-jet}}\,.
\label{eq::nosum}
\end{split}
\end{equation}
Numerical results for $\sigma_{n\text{-jet}}$ with $n=0$ and $n\geq 1$
will be presented in Section~\ref{sec::results}.


\subsection{Cross section with tagged \bld{b}-quarks}
\label{sec:calc:tag}

If realized in Nature, the bottom-quark annihilation process provides a
promising opportunity to measure the bottom-quark Yukawa coupling to the
Higgs boson. To this aim, it is useful to define a proper measurement
function in order to filter events with a specific number of final state
bottom-jets.  As already mentioned, the resulting cross section with one
and two $b$-quarks in the final state has been calculated before through
\nlo{} and \lo{}, respectively, in Ref.~\cite{Campbell:2002zm}. The
calculation is implemented in {\tt MCFM}~\cite{mcfm} which we used in
order to verify our results within our numerical accuracy of $\lesssim
1$\%.

Due to the finite efficiency $\ep_b$ of identifying $b$-jets in the
final state, one needs to distinguish the cross section $\sigma_{nb}$
($n=0,1,2,\ldots$), which is a measure for the number of events with $n$
$b$-jets in the final state, from the cross section
$\sigma_{nb\text{-tag}}$, which concerns the events with $n$ {\it
  tagged} $b$-jets. We require that if the tagging efficiency is 100\%,
both should be the same:\begin{align} \sigma_{nb\text{-tag}}(\ep_b
  \rightarrow 1) = \sigma_{nb}\,.
\end{align}
Therefore, the {\it inclusive} $H+b$-tag cross section at \nlo{}
should be evaluated using~\cite{Campbell:2002zm}
\begin{equation}
\begin{split}
\sigma^\nlo_{\geq1b\text{-tag}} &= \ep_b\sigma^\nlo_{1b} +
\ep_b(2-\ep_b)\sigma^{\lo'}_{2b}\,,
\end{split}
\end{equation}
where $\sigma^{\lo'}_{2b}$ is the exclusive $H+2b$ cross section at
\lo{}, evaluated with \nlo{} \pdf{}s and $\alpha_s$. For the {\it
  exclusive}\footnote{In the context of cross sections with $b$-tags,
  ``exclusive'' only refers to the number of final state $b$-jets; the
  number of jets without $b$-quarks is irrelevant.} 1- and 2-$b$-tagged
cross sections, one may use
\begin{equation}
\begin{split}
\sigma^\nlo_{1b\text{-tag}} &= \ep_b\sigma^\nlo_{1b} +
2\ep_b(1-\ep_b)\sigma^{\lo}_{2b}\,,\\
\sigma^\lo_{2b\text{-tag}} &= \ep_b^2\sigma^\lo_{2b}\,,
\label{eq::12btag}
\end{split}
\end{equation}
where, in contrast to Ref.~\cite{Campbell:2002zm}, we evaluate
$\sigma_{2b}^\lo{}$ consistently with \lo{} \pdf{}s and $\alpha_s$.

The $H+0b$-tag cross section can be calculated at \nnlo{} using
\begin{equation}
\begin{split}
\sigma^\nnlo_{0b\text{-tag}} &= 
\sigma^\nnlo_{0b} +
(1-\ep_b)\sigma^{\nlo}_{1b} +
(1-\ep_b)^2\sigma^{\lo}_{2b}\,.
\label{eq::0btag}
\end{split}
\end{equation}
The first term on the right-hand side refers to events without final
state bottom-quarks, the second and third one concerns events with one
and two final state bottom-quarks, respectively, none of which is
tagged.  $\sigma_{1b}^\nlo$ and $\sigma_{2b}^\lo$ can be calculated
directly with the help of our Monte Carlo program, while the
$0b$-contribution is again obtained by subtracting the inclusive cross
section for $H+b$ production, evaluated with \nnlo{} \pdf{}s and
couplings, from the total rate:
\begin{equation}
\begin{split}
\sigma_{0b}^\nnlo = \sigma_\text{tot}^\nnlo -
\sigma_{\geq1b}^{\nlo'}\,.
\label{eq::sig0b}
\end{split}
\end{equation}

Numerical results for the individual components of
Eqs.\,(\ref{eq::12btag}), (\ref{eq::0btag}), and (\ref{eq::sig0b}) will
be presented in Section\,\ref{sec:results:tag}.



\section{Results}
\label{sec::results}


\subsection{Preliminary remarks}\label{sec:results:prelim}
Before presenting numerical results, let us outline our default
parameters.  Our choice for the central factorization- and
renormalization-scale $\muF$ and $\muR$ is
$\mu_0\equiv\mhiggs/4$. Furthermore, all numbers are produced with the
{\abbrev MSTW2008}~\cite{Martin:2009iq} \pdf{}s which implies that the
numerical value for the strong coupling constant is taken as
$\alpha_s\left(M_Z\right)=0.13939$ at \lo{},
$\alpha_s\left(M_Z\right)=0.12018$ at \nlo{}, and
$\alpha_s\left(M_Z\right)=0.11707$ at \nnlo{}.  The bottom-Higgs Yukawa
coupling is proportional to the bottom-quark mass which we insert in the
$\msbar$-scheme at the scale $\muR$, derived from the input value
$\mbottom(\mbottom)=4.2$\,GeV. No cuts on the Higgs momentum are
applied, and jets are defined using the anti-$k_T$ algorithm with jet
radius: $R=0.4$. A jet is required to have
transverse momentum of $p_{T}^{\text{jet}} > 20\text{ GeV}$ and rapidity
$|y^{\text{jet}}| < 4.8$, unless stated otherwise.

All numbers are evaluated for a \sm{} Higgs boson. An \mssm{} prediction
can be obtained by a proper rescaling of the bottom-quark Yukawa
coupling. This is true for the two {\abbrev CP}-even, and, due to chiral
invariance,\footnote{In the \fs{5}, the bottom quark mass is set to
  zero.}  also for {\abbrev CP}-odd Higgs boson ($h^0$, $H^0$, and $A$).
In this section, we present results for the \lhc{} at 7\,TeV. The
corresponding numbers for 14\,TeV center-of-mass energy can be found in
Appendix~\ref{app::14}.

The theoretical uncertainties of the results presented in this paper
have different sources. One of them arises from the \pdf{}s and the
input value of $\alpha_s(M_Z)$; we evaluate the corresponding
uncertainty  according to the method described in
Ref.~\cite{Martin:2009iq}, using the 68\%{\abbrev CL} \pdf{}s.  Another
source of uncertainty is the truncation of the perturbative series at a
fixed order. It is usually estimated from the dependence of the cross
section on the unphysical scales $\muF$ and $\muR$. In this paper, we
successively fix one of $\muF$ and $\muR$ to $\mu_0$ and vary the other
within $[0.5\,\mu_0,2\,\mu_0]$; the extreme values of the cross section
are then taken as a measure of the perturbative uncertainty. Another
uncertainty is induced by the Monte-Carlo integration; it is small,
however, and will be neglected throughout this paper.

The perturbative uncertainty of the $H+0$-jet cross section has been the
subject of some discussion recently. We will come back to this in
Section~\ref{sec:results:notag}.


\subsection{\bld{H+n}-jet cross sections without jet-flavor tagging}
\label{sec:results:notag}
The first results we present here concern the inclusive $H+$jet and
the exclusive $H+0$-jet (or jet-vetoed) cross sections from bottom-quark
annihilation through \nlo{} and \nnlo{}, respectiveley, where no
flavor-requirement on the outgoing jets is applied.


\begin{figure}
  \begin{center}
    \begin{tabular}{c}
      \subfigure[]{%
        \includegraphics[width=.5\textwidth]{%
          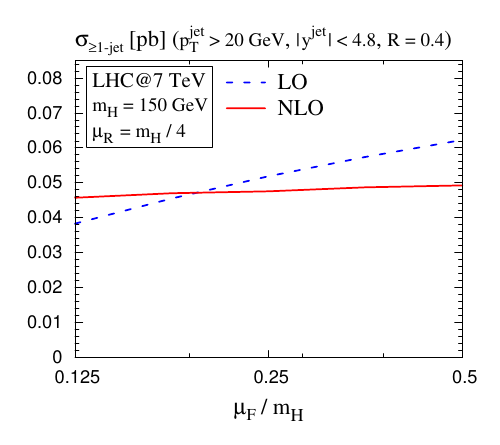}}
      \subfigure[]{%
        \includegraphics[width=.5\textwidth]{%
          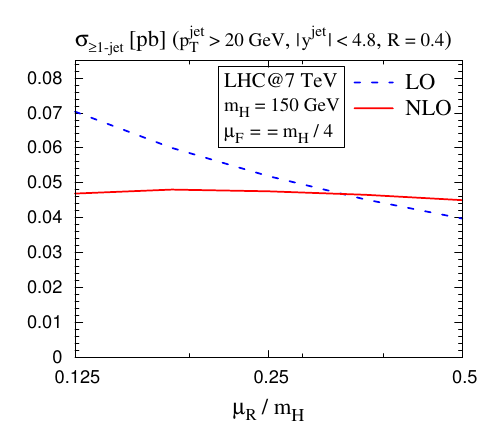}}
    \end{tabular}
    \caption[]{ \label{fig::geq1jet} Scale dependence of the inclusive
      $H+$jet cross section: (a)~$\muF$- and (b)~$\mu_R$-variation.}
  \end{center}
\end{figure}


\begin{figure}
  \begin{center}
    \begin{tabular}{c}
     \subfigure[]{\label{fig:0jet:0jet-muF}%
       \includegraphics[width=.5\textwidth]{%
         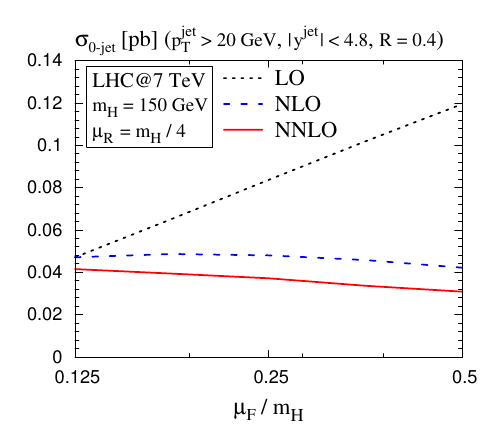}}
      \subfigure[]{\label{fig:0jet:tot-muF}%
      \includegraphics[width=.5\textwidth]{%
        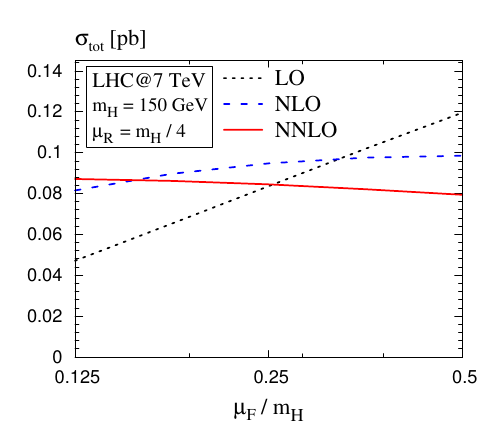}}\\
     \subfigure[]{\label{fig:0jet:0jet-muR}%
       \includegraphics[width=.5\textwidth]{%
         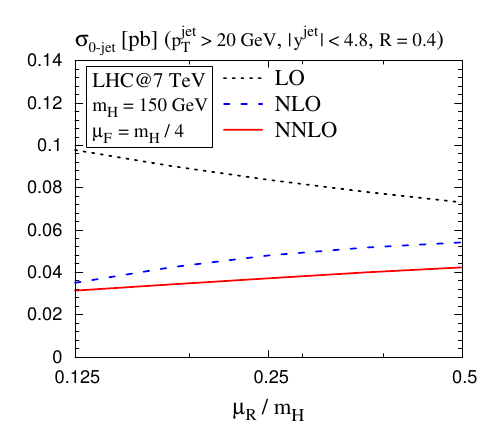}}
      \subfigure[]{\label{fig:0jet:tot-muR}%
        \includegraphics[width=.5\textwidth]{%
          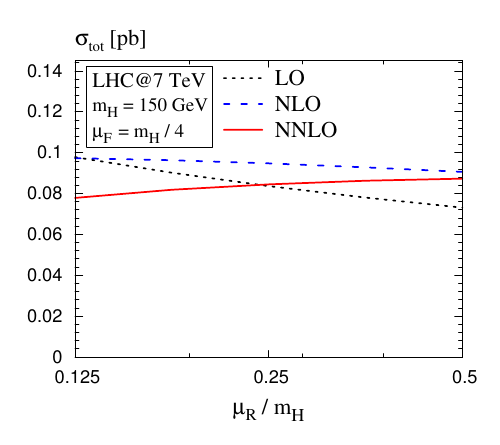}}
    \end{tabular}
    \caption[]{$\muF$-dependence of (a)~the $H+0$-jet cross section and
      (b)~the total cross section, and the $\mu_R$-variation of the same
      quantities in (c) and (d).}
    \label{fig::0jet}
  \end{center}
\end{figure}


\fig{fig::geq1jet} shows the scale variation of the inclusive $H+$jet
rate for a representative Higgs mass of $\mhiggs=150$\,GeV.  One
observes a significant reduction of the scale dependence when going from
\lo{} to \nlo{}, both in $\muF$ and $\muR$, and an excellent consistency
between the \lo{} and the \nlo{} predictions when uncertainties are
taken into account. This justifies our central scale choice of
$\mu_0=\mhiggs/4$.

Concerning the scale variation of the $H+0$-jet cross section, for the
gluon fusion process it has been found to be typically of the same order
of magnitude or even smaller than for the inclusive cross
section\cite{Anastasiou:2009bt,Stewart:2011cf}. It seems unreasonable
then to adopt this variation as the perturbative uncertainty, because
the less inclusive character of this observable is expected to introduce
an additional effect from the truncation of the perturbative series.

However, we do not observe this behavior for bottom-quark annihilation.
For illustration, we consider a Higgs mass of $\mhiggs=150$\,GeV and
compare the scale dependence of the $H+0$-jet to the total cross section
in \fig{fig::0jet}. The leading order curves are identical, since there
are no jets at the partonic level in this case. At higher orders, the
$H+0$-jet cross section is smaller than the total cross section due to
the missing jet contributions, of course. However, the scale dependence
reduces when going to higher orders both for the $H+0$-jet and the total
cross section. While the curves at the various orders for the total
cross section are fairly close to each other in the relevant
$\mu$-range, they move further apart for the $0$-jet bin for the reasons
just discussed. The \nnlo{} curves of the $H+0$-jet and the total cross
section have an almost identical behavior and are just vertically
shifted. The central values differ by roughly a factor of two and, as
opposed to gluon fusion\cite{Anastasiou:2009bt,Stewart:2011cf}, the
relative uncertainty due to scale variation of the $H+0$-jet bin around
$\mu_0$ is more than twice as large as for the total cross section.

For gluon fusion it has been suggested to evaluate the perturbative
uncertainty of the $H+0$-jet cross section from those of its
ingredients, i.e.\ the total and the (primed) inclusive $H+$jet cross
section, see \eqn{eq::0jet}\cite{Stewart:2011cf}. Although we just
argued that for $H$ production in $b\bar b$-annihilation it is not
necessary to adopt this procedure, we provide the corresponding numbers
below for completeness.


\begin{figure}
  \begin{center}
    \begin{tabular}{c}
      \subfigure[]{\label{fig:jetmH:jety48}%
        \includegraphics[width=.5\textwidth]{%
          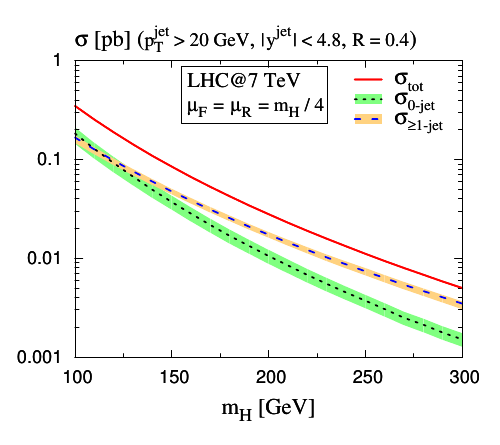}}
      \subfigure[]{\label{fig:jetmH:jety25}%
        \includegraphics[width=.5\textwidth]{%
          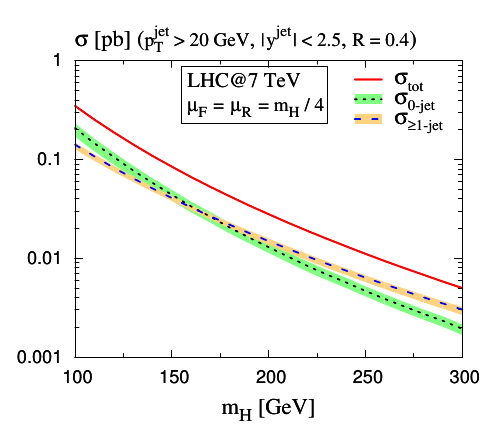}}
    \end{tabular}
    \caption[]{\label{fig::jetmH}Higgs mass dependence of the $H+0$- and
      $\geq1$-jet contributions to the total cross section at \nnlo{}
      and \nlo{}, respectively, for (a)~$|y^{\text{jet}}|<4.8$ and
      (b)~$|y^{\text{jet}}|<2.5$.}
  \end{center}
\end{figure}


\fig{fig::jetmH} shows the decomposition of the total cross section
(solid, red; no uncertainties included) into the exclusive $H+0$-jet
(black, dotted) and the inclusive $H+$jet (blue, dashed) rates. For
illustration, we show the results for both our default jet-rapidity cut
of $|y^\text{jet}|<4.8$ as well as for $|y^\text{jet}|<2.5$. One
observes that the relative contribution of the inclusive $H+$jet rate
increases with the Higgs mass, similar to what was found for the gluon
fusion process~\cite{Catani:2001cr}. The error bands include the
quadratical combination of the perturbative and the \pdf+$\alpha_s$
uncertainties.

The numerical values corresponding to these plots are given in
Appendix~\ref{app::numvals} in Table\,\ref{tab::jety48} and
\ref{tab::jety25} (central values and uncertainties; note
\eqn{eq::nosum}). For completeness, we give in Table\,\ref{tab::tot12p}
also the central values and the perturbative uncertainties of the total
inclusive cross section at \nnlo{} and the primed inclusive $H+$jet
cross section. This provides all numbers required to calculate the scale
uncertainty of the $H+0$-jet cross section from those of its
ingredients~\cite{Stewart:2011cf}.


\subsection{Jet distributions}\label{sec:results:pty}


\begin{figure}[htbp]
  \begin{center}
    \begin{tabular}{c}
     \subfigure[]{%
        \includegraphics[width=.5\textwidth]{%
          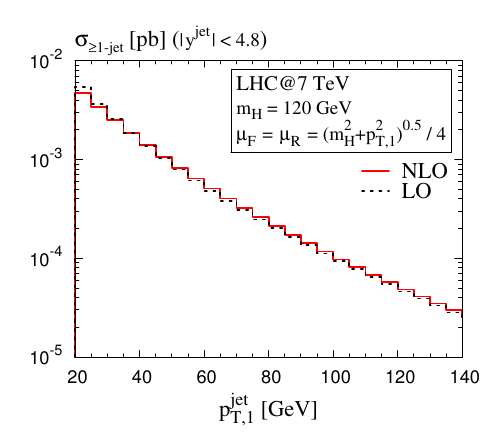}}
      \subfigure[]{%
        \includegraphics[width=.5\textwidth]{%
          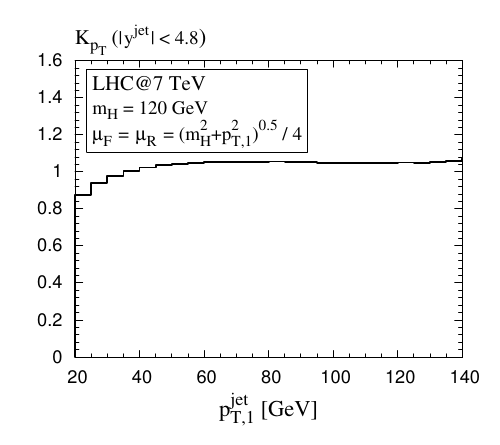}}
    \end{tabular}
    \caption[]{(a)~Transverse momentum distribution of the hardest jet
      in inclusive $H+$jet production, and (b)~the corresponding
      K-factor.}
    \label{fig::pT-dep-y25}
  \end{center}
\end{figure}


\begin{figure}[htbp]
  \begin{center}
    \begin{tabular}{c}
      \subfigure[]{\label{y-cs}\label{fig::jety25}
        \includegraphics[width=.5\textwidth]{%
          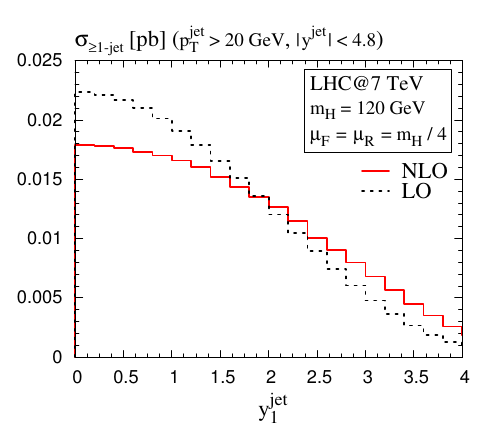}}
     \subfigure[]{\label{y-pT}
     \includegraphics[width=.5\textwidth]{%
         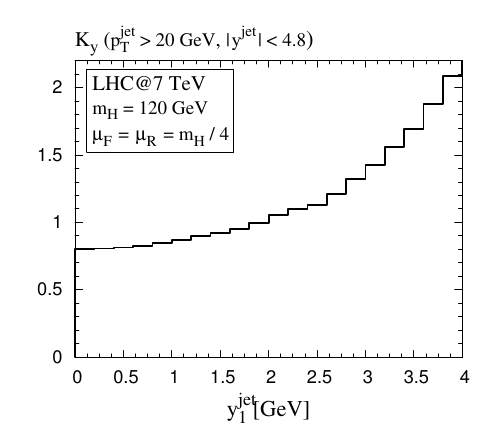}}
    \end{tabular}
    \caption[]{\label{fig::y-dep}(a)~Rapidity distribution of the hardest
      jet in inclusive $H+$jet production and (b)~the corresponding
      K-factor.}
  \end{center}
\end{figure}


Other potentially useful quantities for Higgs studies at the \lhc{} are
the transverse momentum of the hardest jet, $p_{T,1}^\text{jet}$, and
its rapidity $y^\text{jet}_1$. In \fig{fig::pT-dep-y25} we show the
$p_{T,1}^\text{jet}$ distribution for $|y^{\text{jet}}|< 4.8$ of the
inclusive $H+$jet cross section and the corresponding $K$-factor. We
cut the transverse momentum distribution of the hardest jet at
$p_{T,1}^\text{jet}=20$\,GeV, below which resummation is required to
obtain a reliable result. The scale choice
\begin{align}
\mu_F=\mu_R=\frac{1}{4}\sqrt{\mhiggs^2+(p_{T,1}^\text{jet})^2}\,
\end{align}
accounts for possible effects that come from jets with high transverse
momentum. One observes perturbative corrections of typically less than
10\%, and a very weak dependence of the K-factor on $p_{T,1}^\text{jet}$
once it is larger than about 50\,GeV.

Fig.\,\ref{fig::y-dep} shows the rapidity distribution of the hardest
jet. Contrary to the rapidity distribution of the Higgs (see
Ref.~\cite{Harlander:2010cz}), the \nlo{} corrections affect the hardest
jet quite significantly, with the K-factor ranging from 0.8 at central
jet production to about $\text{K}=2$ in the forward- and
backward-region.


\subsection{\bld{H+nb}-jet cross section}\label{sec:results:tag}

In this section, we present numerical results for the $H+nb$-jet cross
sections ($n=0,1,2$) as described in Section~\ref{sec:calc:tag}. Our
default $b$-jet parameters will be chosen according to the corresponding
{\abbrev CMS} analysis: $R=0.5$, $p_T^b > 20$ GeV and
$|y^b| < 2.4$. Results for other parameters are available from the
authors upon request.

Neglecting the bottom-quark mass, as it is required in the \fs{5}, leads
to divergences in the $H+1b$-jet rate which are not cancelled in the sum
of all diagrams and counter terms. They arise in the subprocess $b\bar
b\to b\bar bH$, when a gluon splits into a collinear $b\bar b$ pair. In
Ref.\cite{Campbell:2002zm}, these divergences were circumvented by
introducing an additional cut on the invariant mass of the $b$-quark
pair. We checked that with this cut, these terms contribute less than
1\% to the $H+1b$-jet rate, which is why we neglect them here. The
divergence does not occur in the $H+2b$-jet rate, so all contributions
can safely be included in this case.


\begin{figure}
  \begin{center}
    \begin{tabular}{c}
      \subfigure[]{%
        \includegraphics[width=.5\textwidth]{%
          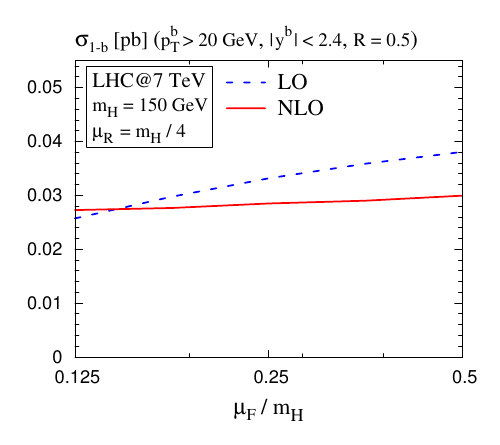}}
      \subfigure[]{%
        \includegraphics[width=.5\textwidth]{%
          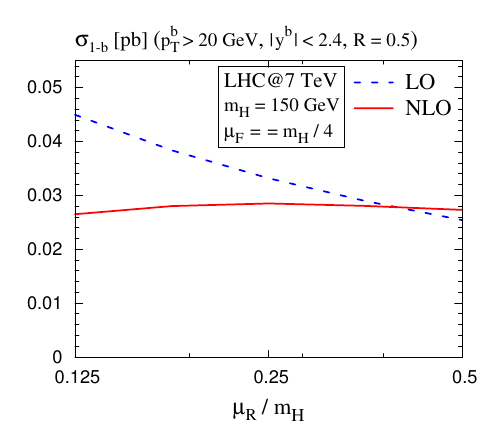}}\\
      \subfigure[]{%
        \includegraphics[width=.5\textwidth]{%
          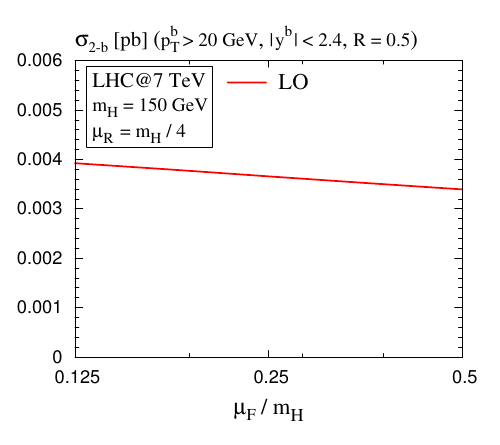}}
      \subfigure[]{\label{fig::2bjet-muR-atl}%
        \includegraphics[width=.5\textwidth]{%
          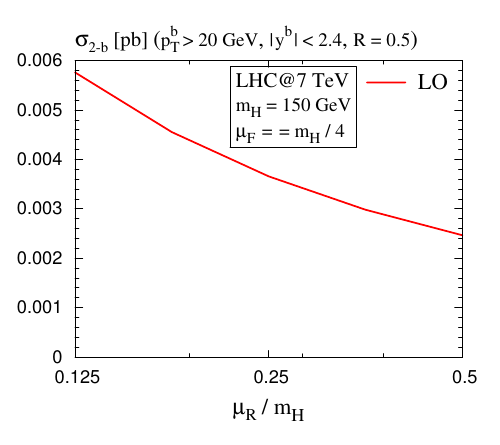}}
    \end{tabular}
    \caption[]{ \label{fig::1bjet} Scale dependence of (a),(b) the
      $H+1b$- and (c),(d) the $H+2b$-jet cross sections $\sigma_{1b}$
      and $\sigma_{2b}$; (a),(c) $\muF$-dependence, (b),(d) $\muR$-dependence.}
  \end{center}
\end{figure}


\begin{figure}
  \begin{center}
    \begin{tabular}{c}
      \subfigure[]{%
        \includegraphics[width=.5\textwidth]{%
          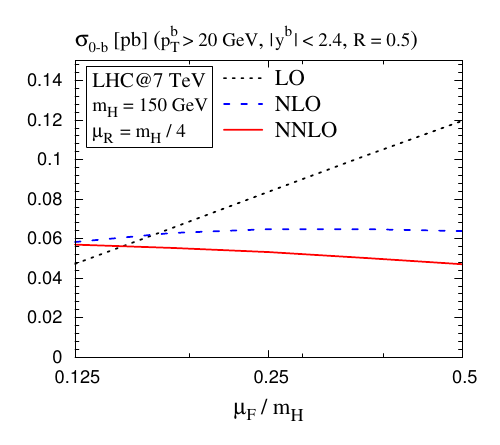}}
      \subfigure[]{%
        \includegraphics[width=.5\textwidth]{%
          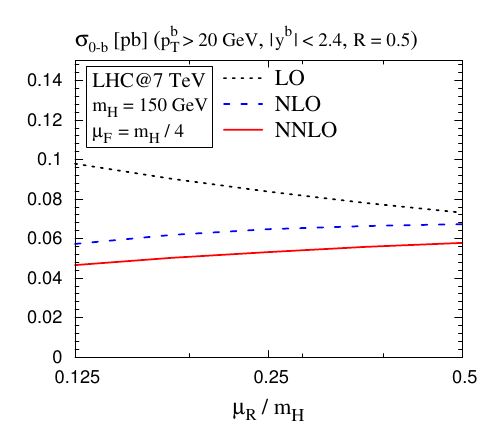}}
    \end{tabular}
    \caption[]{ \label{fig::0bjet} (a) $\muF$- and (b) $\muR$-dependence
      of the $H+0b$-jet cross section.}
  \end{center}
\end{figure}


\begin{figure}
  \begin{center}
    \begin{tabular}{c}
      \subfigure[]{%
       \includegraphics[width=.45\textwidth]{%
         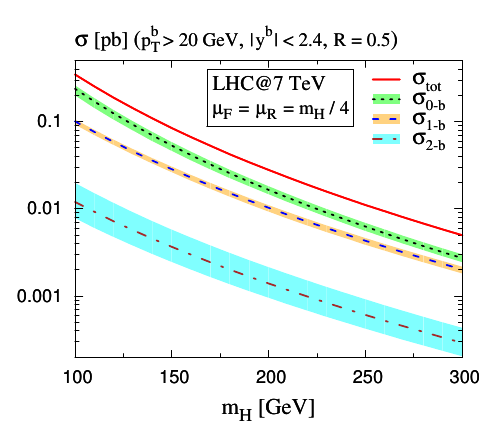}}
      \subfigure[]{%
        \includegraphics[width=.45\textwidth]{%
          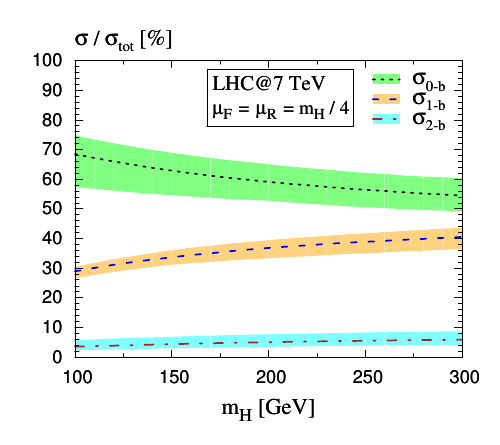}}
    \end{tabular}
    \caption[]{Higgs mass dependence of the $H+nb$-jet contributions
      with respect to the total cross section (a)~in absolut numbers and
      (b)~relative to the total cross section.}
    \label{fig::bjetmH}
  \end{center}
\end{figure}


\fig{fig::1bjet} shows the scale variation of the exclusive $H+1b$-jet
cross section at \lo{} and \nlo{} and the $H+2b$-jet cross section at
\lo{}. Similar to the $H+n$-jet cross sections, one observes a
significant reduction of scale dependence for $\sigma_{1b}$, when going
from \lo{} to \nlo{}. The $H+2b$-jet cross section as a \lo{} quantity
has a larger scale uncertainty which is obviously dominated by the
renormalization scale dependence (see \fig{fig::1bjet}\,(c),(d)).

\fig{fig::0bjet} shows the exclusive $H+0b$-jet cross section at \lo{},
\nlo{} and \nnlo{} as a function of the renormalization/factorization
scale as evaluated in \eqn{eq::sig0b}. We refer to the discussion of
$\sigma_{0\text{-jet}}$ in Section~\ref{sec:results:notag}, since its
qualitative arguments remain valid for the $H+0b$-jet cross section.

Fig.\,\ref{fig::bjetmH} shows the $H+0b$-, $1b$-, and $2b$-cross sections
$\sigma_{0b}^\nnlo$, $\sigma_{1b}^\nlo$, and $\sigma_{2b}^\lo$,
respectively, as they enter the $b$-tagged cross sections according to
Eqs.\,(\ref{eq::12btag}) and (\ref{eq::0btag}). The corresponding
numbers are given in Table\,\ref{tab::012blhc7}. Similar to the
$H+n$-jet cross sections the relative contributions including $b$-jets
increase for higher Higgs masses.



\section{Conclusions}
\label{sec::concl}

The individual contributions of $H+n$-jet events for $n=0$ (jet-veto)
and $n\geq 1$ to the total inclusive cross section for Higgs production
in bottom-quark annihilation have been presented to \nnlo{} and \nlo{}
accuracy. We find a significant reduction of the renormalization- and
factorization scale dependence when going to higher perturbative
orders. In addition, we have studied kinematical distributions for the
hardest emitted jet.

We have also presented an analysis of the $H+nb$-jet rates for $n=0,1,2$,
valid through \nnlo{}, \nlo{}, and \lo{} \qcd{} accuracy, respectively,
and observed a similarly good perturbative convergence as for the
flavor-unspecific jet result. 

The numerical results have been evaluated using realistic
parameters for the \lhc{} experiments and should be directly applicable
to on-going analyses. Results for other parameters can be obtained from
the authors upon request.


\paragraph{Acknowledgements.} We would like to thank the Theory Group at
          {\abbrev CERN}, where most of this work was done, for kind
          hospitality. We are indebted to Fabrice Couderc, Alexander
          Nikitenko, and Markus Schumacher for motivation and
          enlightening discussion, and to John Campbell for
          clarifications concerning {\tt MCFM}. This work was supported
          by {\abbrev BMBF}, contract 05H09PXE.


\def\app#1#2#3{{\it Act.~Phys.~Pol.~}\jref{\bf B #1}{#2}{#3}}
\def\apa#1#2#3{{\it Act.~Phys.~Austr.~}\jref{\bf#1}{#2}{#3}}
\def\annphys#1#2#3{{\it Ann.~Phys.~}\jref{\bf #1}{#2}{#3}}
\def\cmp#1#2#3{{\it Comm.~Math.~Phys.~}\jref{\bf #1}{#2}{#3}}
\def\cpc#1#2#3{{\it Comp.~Phys.~Commun.~}\jref{\bf #1}{#2}{#3}}
\def\epjc#1#2#3{{\it Eur.\ Phys.\ J.\ }\jref{\bf C #1}{#2}{#3}}
\def\fortp#1#2#3{{\it Fortschr.~Phys.~}\jref{\bf#1}{#2}{#3}}
\def\ijmpc#1#2#3{{\it Int.~J.~Mod.~Phys.~}\jref{\bf C #1}{#2}{#3}}
\def\ijmpa#1#2#3{{\it Int.~J.~Mod.~Phys.~}\jref{\bf A #1}{#2}{#3}}
\def\jcp#1#2#3{{\it J.~Comp.~Phys.~}\jref{\bf #1}{#2}{#3}}
\def\jetp#1#2#3{{\it JETP~Lett.~}\jref{\bf #1}{#2}{#3}}
\def\jphysg#1#2#3{{\small\it J.~Phys.~G~}\jref{\bf #1}{#2}{#3}}
\def\jhep#1#2#3{{\small\it JHEP~}\jref{\bf #1}{#2}{#3}}
\def\mpl#1#2#3{{\it Mod.~Phys.~Lett.~}\jref{\bf A #1}{#2}{#3}}
\def\nima#1#2#3{{\it Nucl.~Inst.~Meth.~}\jref{\bf A #1}{#2}{#3}}
\def\npb#1#2#3{{\it Nucl.~Phys.~}\jref{\bf B #1}{#2}{#3}}
\def\nca#1#2#3{{\it Nuovo~Cim.~}\jref{\bf #1A}{#2}{#3}}
\def\plb#1#2#3{{\it Phys.~Lett.~}\jref{\bf B #1}{#2}{#3}}
\def\prc#1#2#3{{\it Phys.~Reports }\jref{\bf #1}{#2}{#3}}
\def\prd#1#2#3{{\it Phys.~Rev.~}\jref{\bf D #1}{#2}{#3}}
\def\pR#1#2#3{{\it Phys.~Rev.~}\jref{\bf #1}{#2}{#3}}
\def\prl#1#2#3{{\it Phys.~Rev.~Lett.~}\jref{\bf #1}{#2}{#3}}
\def\pr#1#2#3{{\it Phys.~Reports }\jref{\bf #1}{#2}{#3}}
\def\ptp#1#2#3{{\it Prog.~Theor.~Phys.~}\jref{\bf #1}{#2}{#3}}
\def\ppnp#1#2#3{{\it Prog.~Part.~Nucl.~Phys.~}\jref{\bf #1}{#2}{#3}}
\def\rmp#1#2#3{{\it Rev.~Mod.~Phys.~}\jref{\bf #1}{#2}{#3}}
\def\sovnp#1#2#3{{\it Sov.~J.~Nucl.~Phys.~}\jref{\bf #1}{#2}{#3}}
\def\sovus#1#2#3{{\it Sov.~Phys.~Usp.~}\jref{\bf #1}{#2}{#3}}
\def\tmf#1#2#3{{\it Teor.~Mat.~Fiz.~}\jref{\bf #1}{#2}{#3}}
\def\tmp#1#2#3{{\it Theor.~Math.~Phys.~}\jref{\bf #1}{#2}{#3}}
\def\yadfiz#1#2#3{{\it Yad.~Fiz.~}\jref{\bf #1}{#2}{#3}}
\def\zpc#1#2#3{{\it Z.~Phys.~}\jref{\bf C #1}{#2}{#3}}
\def\ibid#1#2#3{{ibid.~}\jref{\bf #1}{#2}{#3}}
\def\otherjournal#1#2#3#4{{\it #1}\jref{\bf #2}{#3}{#4}}
\newcommand{\jref}[3]{{\bf #1} (#2) #3}
\newcommand{\hepph}[1]{\href{http://arxiv.org/abs/hep-ph/#1}{\tt [hep-ph/#1]}}
\newcommand{\arxiv}[2]{\href{http://arxiv.org/abs/#1}{\tt [arXiv:#1]}}
\newcommand{\bibentry}[4]{#1, {\it #2}, #3\ifthenelse{\equal{#4}{}}{}{, }#4.}


\newpage
\appendix


\section{Numerical values for the LHC at 7 TeV}
\label{app::numvals}


\begin{table}[h!tbp]
  \centering
      \renewcommand{\arraystretch}{1.2}
\begin{tabular}{|c||c|c|c||c|c|c|}
\hline\multicolumn{7}{|c|}{\lhc{} @ 7\,TeV,\ \ $|y^\text{jet}|<4.8$,\ \ $p_T^\text{jet} > 20$\,GeV,\ \ $R = 0.4$}
\\\hline
$\mhiggs{}$&\multicolumn{1}{|c|}{$\sigma_{0\text{-jet}}^\text{\nnlo}$}&
scale&\pdf{}&
\multicolumn{1}{|c|}{$\sigma_{\geq1\text{-jet}}^{\text{\nlo}}$}&scale&\pdf{}\\[-.4em]

 \footnotesize[GeV]&\multicolumn{1}{|c|}{\footnotesize[fb]}&\footnotesize[\%]&\footnotesize[\%]&\multicolumn{1}{|c|}{\footnotesize[fb]}&\footnotesize[\%]&\footnotesize[\%]\\\hline\hline
$ 100 $& 184 &$ ^{+    14.0}_{-    20.9} $&$^{+     5.1}_{-     4.1}$
&167 &$ ^{+     3.5}_{-     9.9} $&$^{+     3.0}_{-     6.7} $\\
\hline
$ 110 $& 128 &$ ^{+    14.1}_{-    20.0} $&$^{+     5.3}_{-     4.0}$
&126 &$ ^{+     3.3}_{-     9.0} $&$^{+     3.5}_{-     6.2} $\\
\hline
$ 120 $&    92.0 &$ ^{+    14.3}_{-    19.0} $&$^{+     5.4}_{-     3.8}$
&   97.5 &$ ^{+     3.1}_{-     8.2} $&$^{+     4.0}_{-     5.6} $\\
\hline
$ 130 $&    67.0 &$ ^{+    14.4}_{-    18.1} $&$^{+     5.5}_{-     3.7}$
&   75.9 &$ ^{+     3.0}_{-     7.4} $&$^{+     4.5}_{-     5.0} $\\
\hline
$ 140 $&    49.4 &$ ^{+    14.6}_{-    17.1} $&$^{+     5.7}_{-     3.6}$
&   59.7 &$ ^{+     2.8}_{-     6.5} $&$^{+     5.0}_{-     4.4} $\\
\hline
$ 150 $&    37.1 &$ ^{+    14.7}_{-    16.2} $&$^{+     5.8}_{-     3.4}$
&   47.4 &$ ^{+     2.6}_{-     5.7} $&$^{+     5.5}_{-     3.9} $\\
\hline
$ 160 $&    28.2 &$ ^{+    14.6}_{-    16.1} $&$^{+     5.7}_{-     3.8}$
&   38.3 &$ ^{+     2.6}_{-     5.7} $&$^{+     5.6}_{-     3.8} $\\
\hline
$ 170 $&    21.6 &$ ^{+    14.5}_{-    16.0} $&$^{+     5.6}_{-     4.2}$
&   31.0 &$ ^{+     2.6}_{-     5.7} $&$^{+     5.6}_{-     3.8} $\\
\hline
$ 180 $&    16.8 &$ ^{+    14.4}_{-    15.9} $&$^{+     5.4}_{-     4.6}$
&   25.4 &$ ^{+     2.5}_{-     5.6} $&$^{+     5.7}_{-     3.8} $\\
\hline
$ 190 $&    13.3 &$ ^{+    14.3}_{-    15.9} $&$^{+     5.3}_{-     4.9}$
&   20.9 &$ ^{+     2.5}_{-     5.6} $&$^{+     5.7}_{-     3.7} $\\
\hline
$ 200 $&    10.5 &$ ^{+    14.2}_{-    15.8} $&$^{+     5.2}_{-     5.3}$
&   17.3 &$ ^{+     2.5}_{-     5.6} $&$^{+     5.8}_{-     3.7} $\\
\hline
$ 210 $&    8.43 &$ ^{+    14.1}_{-    15.5} $&$^{+     5.2}_{-     5.7}$
&   14.4 &$ ^{+     2.5}_{-     6.1} $&$^{+     6.1}_{-     3.8} $\\
\hline
$ 220 $&    6.75 &$ ^{+    14.0}_{-    15.2} $&$^{+     5.2}_{-     6.1}$
&   12.1 &$ ^{+     2.5}_{-     6.5} $&$^{+     6.4}_{-     4.0} $\\
\hline
$ 230 $&    5.49 &$ ^{+    13.9}_{-    14.9} $&$^{+     5.3}_{-     6.4}$
&   10.2 &$ ^{+     2.5}_{-     7.0} $&$^{+     6.8}_{-     4.2} $\\
\hline
$ 240 $&    4.49 &$ ^{+    13.8}_{-    14.6} $&$^{+     5.3}_{-     6.8}$
&   8.64 &$ ^{+     2.5}_{-     7.4} $&$^{+     7.1}_{-     4.3} $\\
\hline
$ 250 $&    3.71 &$ ^{+    13.7}_{-    14.2} $&$^{+     5.3}_{-     7.2}$
&   7.36 &$ ^{+     2.5}_{-     7.9} $&$^{+     7.4}_{-     4.5} $\\
\hline
$ 260 $&    3.04 &$ ^{+    13.6}_{-    14.4} $&$^{+     6.1}_{-     6.7}$
&   6.28 &$ ^{+     2.5}_{-     8.1} $&$^{+     6.9}_{-     5.3} $\\
\hline
$ 270 $&    2.49 &$ ^{+    13.4}_{-    14.5} $&$^{+     6.8}_{-     6.2}$
&   5.38 &$ ^{+     2.6}_{-     8.3} $&$^{+     6.3}_{-     6.0} $\\
\hline
$ 280 $&    2.11 &$ ^{+    13.3}_{-    14.7} $&$^{+     7.6}_{-     5.8}$
&   4.64 &$ ^{+     2.7}_{-     8.5} $&$^{+     5.8}_{-     6.8} $\\
\hline
$ 290 $&    1.77 &$ ^{+    13.2}_{-    14.8} $&$^{+     8.4}_{-     5.3}$
&   4.01 &$ ^{+     2.8}_{-     8.6} $&$^{+     5.2}_{-     7.6} $\\
\hline
$ 300 $&    1.50 &$ ^{+    13.1}_{-    14.9} $&$^{+     9.1}_{-     4.8}$
&   3.47 &$ ^{+     2.8}_{-     8.8} $&$^{+     4.7}_{-     8.4} $\\
\hline
\end{tabular}

      \caption[]{\label{tab::jety48}Central values, scale and \pdf{}
        uncertainties for the exclusive $H+0$-jet (\nnlo) and inclusive
        $H+$jet (\nlo) cross section for $|y^{\text{jet}}| < 4.8$, see
        also \fig{fig::jetmH}\,(a).}
\end{table}


\begin{table}[h!tbp]
  \centering
      \renewcommand{\arraystretch}{1.2}
\begin{tabular}{|c||c|c|c||c|c|c|}
\hline\multicolumn{7}{|c|}{\lhc{} @ 7\,TeV,\ \ $|y^\text{jet}|<2.5$,\ \ $p_T^\text{jet} > 20$\,GeV,\ \ $R = 0.4$}
\\\hline
$\mhiggs{}$&\multicolumn{1}{|c|}{$\sigma_{0\text{-jet}}^\text{\nnlo}$}&
scale&\pdf{}&
\multicolumn{1}{|c|}{$\sigma_{\geq1\text{-jet}}^{\text{\nlo}}$}&scale&\pdf{}\\[-.4em]

 \footnotesize[GeV]&\multicolumn{1}{|c|}{\footnotesize[fb]}&\footnotesize[\%]&\footnotesize[\%]&\multicolumn{1}{|c|}{\footnotesize[fb]}&\footnotesize[\%]&\footnotesize[\%]\\\hline\hline
$ 100 $& 209 &$ ^{+    10.9}_{-    17.7} $&$^{+     3.2}_{-     3.1}$
&141 &$ ^{+     3.6}_{-     9.2} $&$^{+     3.7}_{-     4.6} $\\
\hline
$ 110 $& 146 &$ ^{+    10.9}_{-    16.8} $&$^{+     3.5}_{-     3.0}$
&107 &$ ^{+     3.0}_{-     8.4} $&$^{+     4.0}_{-     4.4} $\\
\hline
$ 120 $& 106 &$ ^{+    10.9}_{-    15.9} $&$^{+     3.8}_{-     2.9}$
&   82.5 &$ ^{+     2.4}_{-     7.6} $&$^{+     4.2}_{-     4.2} $\\
\hline
$ 130 $&    77.4 &$ ^{+    10.9}_{-    14.9} $&$^{+     4.1}_{-     2.9}$
&   64.5 &$ ^{+     1.8}_{-     6.8} $&$^{+     4.5}_{-     4.0} $\\
\hline
$ 140 $&    57.9 &$ ^{+    10.9}_{-    14.0} $&$^{+     4.4}_{-     2.8}$
&   50.9 &$ ^{+     1.2}_{-     6.0} $&$^{+     4.7}_{-     3.8} $\\
\hline
$ 150 $&    44.0 &$ ^{+    10.9}_{-    13.1} $&$^{+     4.7}_{-     2.7}$
&   40.7 &$ ^{+     0.5}_{-     5.2} $&$^{+     5.0}_{-     3.6} $\\
\hline
$ 160 $&    33.5 &$ ^{+    10.8}_{-    12.7} $&$^{+     4.5}_{-     3.3}$
&   32.6 &$ ^{+     0.8}_{-     5.6} $&$^{+     5.2}_{-     3.6} $\\
\hline
$ 170 $&    26.2 &$ ^{+    10.7}_{-    12.3} $&$^{+     4.3}_{-     3.9}$
&   26.6 &$ ^{+     1.0}_{-     6.0} $&$^{+     5.4}_{-     3.7} $\\
\hline
$ 180 $&    20.4 &$ ^{+    10.6}_{-    11.9} $&$^{+     4.1}_{-     4.6}$
&   21.8 &$ ^{+     1.2}_{-     6.4} $&$^{+     5.6}_{-     3.8} $\\
\hline
$ 190 $&    16.1 &$ ^{+    10.5}_{-    11.5} $&$^{+     3.9}_{-     5.2}$
&   17.9 &$ ^{+     1.5}_{-     6.7} $&$^{+     5.8}_{-     3.9} $\\
\hline
$ 200 $&    12.9 &$ ^{+    10.4}_{-    11.1} $&$^{+     3.6}_{-     5.8}$
&   14.9 &$ ^{+     1.7}_{-     7.1} $&$^{+     6.1}_{-     3.9} $\\
\hline
$ 210 $&    10.4 &$ ^{+    10.2}_{-    11.0} $&$^{+     4.2}_{-     5.8}$
&   12.4 &$ ^{+     1.7}_{-     7.5} $&$^{+     5.9}_{-     4.1} $\\
\hline
$ 220 $&    8.34 &$ ^{+    10.0}_{-    10.9} $&$^{+     4.8}_{-     5.7}$
&   10.4 &$ ^{+     1.7}_{-     7.8} $&$^{+     5.6}_{-     4.3} $\\
\hline
$ 230 $&    6.84 &$ ^{+     9.8}_{-    10.8} $&$^{+     5.4}_{-     5.7}$
&   8.81 &$ ^{+     1.8}_{-     8.1} $&$^{+     5.4}_{-     4.5} $\\
\hline
$ 240 $&    5.68 &$ ^{+     9.6}_{-    10.7} $&$^{+     6.0}_{-     5.6}$
&   7.49 &$ ^{+     1.8}_{-     8.4} $&$^{+     5.2}_{-     4.7} $\\
\hline
$ 250 $&    4.66 &$ ^{+     9.3}_{-    10.6} $&$^{+     6.6}_{-     5.6}$
&   6.37 &$ ^{+     1.8}_{-     8.7} $&$^{+     5.0}_{-     4.9} $\\
\hline
$ 260 $&    3.87 &$ ^{+     9.3}_{-    10.5} $&$^{+     6.4}_{-     6.1}$
&   5.44 &$ ^{+     2.0}_{-     9.0} $&$^{+     5.1}_{-     5.2} $\\
\hline
$ 270 $&    3.25 &$ ^{+     9.3}_{-    10.5} $&$^{+     6.1}_{-     6.7}$
&   4.67 &$ ^{+     2.1}_{-     9.2} $&$^{+     5.2}_{-     5.5} $\\
\hline
$ 280 $&    2.72 &$ ^{+     9.3}_{-    10.4} $&$^{+     5.9}_{-     7.2}$
&   4.01 &$ ^{+     2.3}_{-     9.5} $&$^{+     5.2}_{-     5.8} $\\
\hline
$ 290 $&    2.32 &$ ^{+     9.2}_{-    10.4} $&$^{+     5.7}_{-     7.7}$
&   3.48 &$ ^{+     2.5}_{-     9.7} $&$^{+     5.3}_{-     6.1} $\\
\hline
$ 300 $&    1.93 &$ ^{+     9.2}_{-    10.3} $&$^{+     5.4}_{-     8.3}$
&   3.02 &$ ^{+     2.6}_{-    10.0} $&$^{+     5.4}_{-     6.4} $\\
\hline
\end{tabular}

      \caption[]{ \label{tab::jety25}Same as Table\,\ref{tab::jety48},
        but for $|y^{\text{jet}}| < 2.5$, see also
        \fig{fig::jetmH}\,(b).}
\end{table}


\begin{table}[h!tbp]
  \centering
      \renewcommand{\arraystretch}{1.2}
\begin{tabular}{|c||c|c||c|c||c|c|}
\hline\multicolumn{7}{|c|}{\lhc{} @ 7\,TeV}
\\\hline
&&&
\multicolumn{4}{c|}{$p_T^\text{jet}>20$\,GeV,\ \ $R = 0.4$}\\
\cline{4-7}
&&&
\multicolumn{2}{c||}{$|y^\text{jet}|<4.8$}
&
\multicolumn{2}{c|}{$|y^\text{jet}|<2.5$}
\\\cline{4-7}
$\mhiggs{}$
&
\multicolumn{1}{c|}{$\sigma_\text{tot}^\text{\nnlo}$}
&
scale
&
\multicolumn{1}{c|}{$\sigma_{\geq1\text{-jet}}^{\text{\nlo}'}$}
&
scale
&
\multicolumn{1}{c|}{$\sigma_{\geq1\text{-jet}}^{\text{\nlo}'}$}
&
scale
\\[-.4em]

 \footnotesize[GeV]
&
\multicolumn{1}{c|}{\footnotesize[fb]}
&
\footnotesize[\%]
&
\multicolumn{1}{c|}{\footnotesize[fb]}
&
\footnotesize[\%]
&
\multicolumn{1}{c|}{\footnotesize[fb]}
&
\footnotesize[\%]
\\\hline\hline
$ 100 $& 349 &$^{+     2.9}_{-    10.0}$&
165&$^{+     2.1}_{-     9.4}$&140&$^{+     3.4}_{-     9.0}$\\\hline
$ 110 $& 254 &$^{+     3.0}_{-     9.6}$&
126&$^{+     2.1}_{-     8.6}$&108&$^{+     3.0}_{-     8.2}$\\\hline
$ 120 $& 188 &$^{+     3.1}_{-     9.1}$&
   96.3&$^{+     2.0}_{-     7.9}$&   82.4&$^{+     2.7}_{-     7.4}$\\\hline
$ 130 $& 142 &$^{+     3.1}_{-     8.7}$&
   75.1&$^{+     2.0}_{-     7.2}$&   64.7&$^{+     2.3}_{-     6.6}$\\\hline
$ 140 $& 109 &$^{+     3.2}_{-     8.3}$&
   59.4&$^{+     1.9}_{-     6.4}$&   50.9&$^{+     1.9}_{-     5.8}$\\\hline
$ 150 $&    84.5 &$^{+     3.2}_{-     7.8}$&
   47.4&$^{+     1.9}_{-     5.7}$&   40.5&$^{+     1.5}_{-     5.1}$\\\hline
$ 160 $&    66.4 &$^{+     3.2}_{-     7.6}$&
   38.2&$^{+     1.7}_{-     5.5}$&   32.9&$^{+     1.6}_{-     5.3}$\\\hline
$ 170 $&    52.7 &$^{+     3.2}_{-     7.5}$&
   31.1&$^{+     1.6}_{-     5.3}$&   26.5&$^{+     1.6}_{-     5.5}$\\\hline
$ 180 $&    42.3 &$^{+     3.2}_{-     7.3}$&
   25.4&$^{+     1.4}_{-     5.1}$&   21.9&$^{+     1.6}_{-     5.7}$\\\hline
$ 190 $&    34.2 &$^{+     3.2}_{-     7.1}$&
   20.9&$^{+     1.3}_{-     4.9}$&   18.1&$^{+     1.7}_{-     6.0}$\\\hline
$ 200 $&    27.9 &$^{+     3.2}_{-     6.9}$&
   17.4&$^{+     1.1}_{-     4.8}$&   15.0&$^{+     1.7}_{-     6.2}$\\\hline
$ 210 $&    22.9 &$^{+     3.2}_{-     6.8}$&
   14.5&$^{+     1.4}_{-     5.2}$&   12.5&$^{+     1.8}_{-     6.4}$\\\hline
$ 220 $&    18.9 &$^{+     3.2}_{-     6.7}$&
   12.2&$^{+     1.7}_{-     5.7}$&   10.6&$^{+     1.9}_{-     6.7}$\\\hline
$ 230 $&    15.7 &$^{+     3.1}_{-     6.5}$&
   10.2&$^{+     1.9}_{-     6.1}$&   8.90&$^{+     2.1}_{-     6.9}$\\\hline
$ 240 $&    13.2 &$^{+     3.1}_{-     6.4}$&
   8.68&$^{+     2.2}_{-     6.6}$&   7.49&$^{+     2.2}_{-     7.2}$\\\hline
$ 250 $&    11.1 &$^{+     3.1}_{-     6.3}$&
   7.37&$^{+     2.4}_{-     7.1}$&   6.42&$^{+     2.3}_{-     7.4}$\\\hline
$ 260 $&    9.36 &$^{+     3.1}_{-     6.2}$&
   6.31&$^{+     2.5}_{-     7.3}$&   5.49&$^{+     2.3}_{-     7.7}$\\\hline
$ 270 $&    7.95 &$^{+     3.1}_{-     6.1}$&
   5.46&$^{+     2.6}_{-     7.6}$&   4.70&$^{+     2.4}_{-     8.0}$\\\hline
$ 280 $&    6.78 &$^{+     3.0}_{-     5.9}$&
   4.67&$^{+     2.7}_{-     7.8}$&   4.05&$^{+     2.4}_{-     8.4}$\\\hline
$ 290 $&    5.80 &$^{+     3.0}_{-     5.8}$&
   4.03&$^{+     2.8}_{-     8.1}$&   3.48&$^{+     2.5}_{-     8.7}$\\\hline
$ 300 $&    4.98 &$^{+     3.0}_{-     5.7}$&
   3.48&$^{+     2.9}_{-     8.3}$&   3.06&$^{+     2.5}_{-     9.0}$\\\hline
\end{tabular}

	  \caption[]{\label{tab::tot12p}Total cross section
            $\sigma_\text{tot}^\nnlo$, and
            $\sigma_{\geq1\text{-jet}}^{\nlo'}$ for two different jet
            rapidity cuts, including the associated scale uncertainties.}
\end{table}


\begin{table}[h!tbp]
  \centering
      \renewcommand{\arraystretch}{1.2}
\begin{tabular}{|c||c|c|c||c|c|c||c|c|c|}
\hline\multicolumn{10}{|c|}{\lhc{} @ 7\,TeV,\ \ $|y^{b}|<2.4$,\ \ $p_T^b > 20$\,GeV,\ \ $R = 0.5$}
\\\hline
$\mhiggs{}$&\multicolumn{1}{|c|}{$\sigma_{0b}^\text{\nnlo{}}$}&
scale&
\pdf{}&
\multicolumn{1}{|c|}{$\sigma_{1b}^\text{\nlo}$}&
scale&
\pdf{}&
\multicolumn{1}{|c|}{$\sigma_{2b}^\text{\lo}$}&
scale&
\pdf{}\\[-.4em]
\footnotesize[GeV]&\multicolumn{1}{|c|}{\footnotesize[fb]}&\footnotesize[\%]&\footnotesize[\%]&\multicolumn{1}{|c|}{\footnotesize[fb]}&\footnotesize[\%]&\footnotesize[\%]&\multicolumn{1}{|c|}{\footnotesize[fb]}&\footnotesize[\%]&\footnotesize[\%]\\\hline\hline
100&239&$^{+     8.3}_{-    15.9} $&$^{+     3.8}_{-     3.2}$&
101 &$ ^{+     4.1}_{-     7.0} $&$^{+     3.4}_{-     3.8}$&
   12.0 &$ ^{+    64.7}_{-    34.8} $&$^{+     2.5}_{-     2.6}$
 \\
\hline
110&170&$^{+     8.5}_{-    15.2} $&$^{+     3.8}_{-     3.5}$&
   76.3 &$ ^{+     4.2}_{-     7.0} $&$^{+     3.8}_{-     3.6}$&
   9.24 &$ ^{+    63.3}_{-    34.4} $&$^{+     2.4}_{-     2.7}$
 \\
\hline
120&124&$^{+     8.6}_{-    14.4} $&$^{+     3.7}_{-     3.9}$&
   58.6 &$ ^{+     4.3}_{-     7.0} $&$^{+     4.2}_{-     3.5}$&
   7.20 &$ ^{+    61.9}_{-    34.0} $&$^{+     2.4}_{-     2.8}$
 \\
\hline
130&   92.2&$^{+     8.7}_{-    13.7} $&$^{+     3.7}_{-     4.2}$&
   45.5 &$ ^{+     4.3}_{-     6.9} $&$^{+     4.6}_{-     3.4}$&
   5.69 &$ ^{+    60.5}_{-    33.5} $&$^{+     2.4}_{-     2.9}$
 \\
\hline
140&   69.5&$^{+     8.8}_{-    12.9} $&$^{+     3.6}_{-     4.6}$&
   35.8 &$ ^{+     4.4}_{-     6.9} $&$^{+     5.0}_{-     3.2}$&
   4.54 &$ ^{+    59.0}_{-    33.1} $&$^{+     2.3}_{-     3.0}$
 \\
\hline
150&   53.1&$^{+     9.0}_{-    12.2} $&$^{+     3.6}_{-     4.9}$&
   28.4 &$ ^{+     4.5}_{-     6.9} $&$^{+     5.4}_{-     3.1}$&
   3.66 &$ ^{+    57.6}_{-    32.7} $&$^{+     2.3}_{-     3.1}$
 \\
\hline
160&   41.2&$^{+     9.0}_{-    11.7} $&$^{+     3.7}_{-     4.8}$&
   22.9 &$ ^{+     4.7}_{-     7.2} $&$^{+     5.4}_{-     3.4}$&
   2.97 &$ ^{+    56.8}_{-    32.4} $&$^{+     2.4}_{-     3.1}$
 \\
\hline
170&   32.3&$^{+     9.0}_{-    11.3} $&$^{+     3.9}_{-     4.6}$&
   18.5 &$ ^{+     4.8}_{-     7.5} $&$^{+     5.3}_{-     3.6}$&
   2.44 &$ ^{+    56.0}_{-    32.1} $&$^{+     2.4}_{-     3.1}$
 \\
\hline
180&   25.6&$^{+     9.1}_{-    10.8} $&$^{+     4.0}_{-     4.5}$&
   15.1 &$ ^{+     4.9}_{-     7.9} $&$^{+     5.2}_{-     3.9}$&
   2.01 &$ ^{+    55.2}_{-    31.8} $&$^{+     2.5}_{-     3.1}$
 \\
\hline
190&   20.5&$^{+     9.1}_{-    10.4} $&$^{+     4.1}_{-     4.4}$&
   12.4 &$ ^{+     5.0}_{-     8.2} $&$^{+     5.1}_{-     4.1}$&
   1.67 &$ ^{+    54.3}_{-    31.6} $&$^{+     2.6}_{-     3.2}$
 \\
\hline
200&   16.5&$^{+     9.1}_{-     9.9} $&$^{+     4.3}_{-     4.2}$&
   10.3 &$ ^{+     5.1}_{-     8.5} $&$^{+     5.1}_{-     4.4}$&
   1.39 &$ ^{+    53.5}_{-    31.3} $&$^{+     2.6}_{-     3.2}$
 \\
\hline
210&   13.4&$^{+     9.1}_{-     9.8} $&$^{+     4.3}_{-     4.3}$&
   8.53 &$ ^{+     5.2}_{-     8.5} $&$^{+     5.0}_{-     4.5}$&
   1.17 &$ ^{+    52.9}_{-    31.1} $&$^{+     2.8}_{-     3.2}$
 \\
\hline
220&   11.0&$^{+     9.1}_{-     9.7} $&$^{+     4.4}_{-     4.4}$&
   7.13 &$ ^{+     5.3}_{-     8.5} $&$^{+     5.0}_{-     4.6}$&
  0.985 &$ ^{+    52.4}_{-    30.9} $&$^{+     2.9}_{-     3.2}$
 \\
\hline
230&   9.04&$^{+     9.0}_{-     9.6} $&$^{+     4.5}_{-     4.5}$&
   6.01 &$ ^{+     5.3}_{-     8.6} $&$^{+     5.0}_{-     4.7}$&
  0.835 &$ ^{+    51.8}_{-    30.7} $&$^{+     3.1}_{-     3.2}$
 \\
\hline
240&   7.49&$^{+     9.0}_{-     9.5} $&$^{+     4.6}_{-     4.6}$&
   5.08 &$ ^{+     5.4}_{-     8.6} $&$^{+     4.9}_{-     4.9}$&
  0.710 &$ ^{+    51.3}_{-    30.5} $&$^{+     3.2}_{-     3.2}$
 \\
\hline
250&   6.25&$^{+     9.0}_{-     9.4} $&$^{+     4.7}_{-     4.7}$&
   4.30 &$ ^{+     5.4}_{-     8.6} $&$^{+     4.9}_{-     5.0}$&
  0.607 &$ ^{+    50.7}_{-    30.3} $&$^{+     3.4}_{-     3.2}$
 \\
\hline
260&   5.24&$^{+     9.0}_{-     9.2} $&$^{+     4.9}_{-     4.9}$&
   3.67 &$ ^{+     5.5}_{-     8.7} $&$^{+     5.0}_{-     4.9}$&
  0.520 &$ ^{+    50.3}_{-    30.1} $&$^{+     3.4}_{-     3.3}$
 \\
\hline
270&   4.41&$^{+     9.0}_{-     8.9} $&$^{+     5.1}_{-     5.0}$&
   3.14 &$ ^{+     5.5}_{-     8.7} $&$^{+     5.0}_{-     4.9}$&
  0.448 &$ ^{+    49.8}_{-    30.0} $&$^{+     3.4}_{-     3.4}$
 \\
\hline
280&   3.74&$^{+     9.0}_{-     8.7} $&$^{+     5.3}_{-     5.1}$&
   2.70 &$ ^{+     5.6}_{-     8.7} $&$^{+     5.1}_{-     4.8}$&
  0.386 &$ ^{+    49.4}_{-    29.8} $&$^{+     3.5}_{-     3.4}$
 \\
\hline
290&   3.18&$^{+     8.9}_{-     8.5} $&$^{+     5.5}_{-     5.3}$&
   2.33 &$ ^{+     5.6}_{-     8.8} $&$^{+     5.2}_{-     4.8}$&
  0.335 &$ ^{+    49.0}_{-    29.7} $&$^{+     3.5}_{-     3.5}$
 \\
\hline
300&   2.71&$^{+     8.9}_{-     8.2} $&$^{+     5.7}_{-     5.4}$&
   2.02 &$ ^{+     5.7}_{-     8.8} $&$^{+     5.3}_{-     4.7}$&
  0.291 &$ ^{+    48.6}_{-    29.5} $&$^{+     3.5}_{-     3.5}$
 \\
\hline
\end{tabular}

	  \caption{\label{tab::012blhc7}Central values, scale and \pdf{}
            uncertainties for the exclusive $H+nb$-jet ($n=0,1,2$) cross
            section at \nnlo{}, \nlo{}, and \lo{}, respectively. See
            also \fig{fig::bjetmH}.}
\end{table}


\newpage

\section{Results for the LHC at 14 TeV}
\label{app::14}


\begin{figure}[h!tb]
  \begin{center}
    \begin{tabular}{c}
      \subfigure[]{\label{fig:jetmH:S14jety48}%
        \includegraphics[width=.5\textwidth]{%
          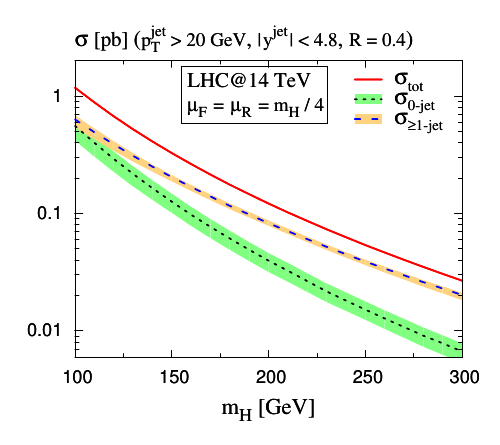}}
      \subfigure[]{\label{fig:jetmH:S14jety25}%
        \includegraphics[width=.5\textwidth]{%
          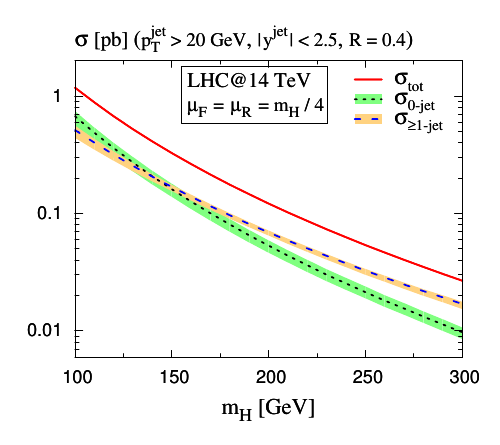}}
    \end{tabular}
    \caption[]{Higgs mass dependence of the $H+0$- and $\geq1$-jet
      contributions to the total cross section at \nnlo{} and \nlo{},
      respectively, for (a) $|y^{\text{jet}}|<4.8$ and (b)
      $|y^{\text{jet}}|<2.5$.}
    \label{fig::jetmH-14}
  \end{center}
\end{figure}


\begin{table}[h!tbp]
  \centering
      \renewcommand{\arraystretch}{1.2}
\begin{tabular}{|c||c|c|c||c|c|c|}
\hline\multicolumn{7}{|c|}{\lhc{} @ 14\,TeV,\ \ $|y^\text{jet}|<4.8$,\ \ $p_T^\text{jet} > 20$\,GeV,\ \ $R = 0.4$}
\\\hline
$\mhiggs{}$&\multicolumn{1}{|c|}{$\sigma_{0\text{-jet}}^\text{\nnlo}$}&
scale&\pdf{}&
\multicolumn{1}{|c|}{$\sigma_{\geq1\text{-jet}}^{\text{\nlo}}$}&scale&\pdf{}\\[-.4em]

 \footnotesize[GeV]&\multicolumn{1}{|c|}{\footnotesize[fb]}&\footnotesize[\%]&\footnotesize[\%]&\multicolumn{1}{|c|}{\footnotesize[fb]}&\footnotesize[\%]&\footnotesize[\%]\\\hline\hline
$ 100 $& 558 &$ ^{+    14.4}_{-    24.6} $&$^{+     3.2}_{-     3.1}$
&632 &$ ^{+     8.3}_{-    13.3} $&$^{+     2.6}_{-     4.6} $\\
\hline
$ 110 $& 400 &$ ^{+    14.9}_{-    23.4} $&$^{+     3.3}_{-     3.7}$
&492 &$ ^{+     7.2}_{-    12.2} $&$^{+     2.7}_{-     4.3} $\\
\hline
$ 120 $& 292 &$ ^{+    15.4}_{-    22.3} $&$^{+     3.3}_{-     4.2}$
&388 &$ ^{+     6.0}_{-    11.1} $&$^{+     2.7}_{-     4.0} $\\
\hline
$ 130 $& 217 &$ ^{+    15.9}_{-    21.1} $&$^{+     3.4}_{-     4.7}$
&310 &$ ^{+     4.9}_{-     9.9} $&$^{+     2.7}_{-     3.7} $\\
\hline
$ 140 $& 165 &$ ^{+    16.3}_{-    19.9} $&$^{+     3.5}_{-     5.2}$
&251 &$ ^{+     3.8}_{-     8.8} $&$^{+     2.7}_{-     3.4} $\\
\hline
$ 150 $& 126 &$ ^{+    16.8}_{-    18.8} $&$^{+     3.5}_{-     5.7}$
&204 &$ ^{+     2.6}_{-     7.6} $&$^{+     2.8}_{-     3.1} $\\
\hline
$ 160 $&    97.7 &$ ^{+    16.6}_{-    18.7} $&$^{+     3.2}_{-     6.1}$
&168 &$ ^{+     2.4}_{-     7.0} $&$^{+     3.3}_{-     3.0} $\\
\hline
$ 170 $&    76.8 &$ ^{+    16.5}_{-    18.6} $&$^{+     2.8}_{-     6.6}$
&139 &$ ^{+     2.3}_{-     6.4} $&$^{+     3.8}_{-     2.8} $\\
\hline
$ 180 $&    61.2 &$ ^{+    16.3}_{-    18.6} $&$^{+     2.5}_{-     7.0}$
&116 &$ ^{+     2.1}_{-     5.7} $&$^{+     4.2}_{-     2.7} $\\
\hline
$ 190 $&    49.0 &$ ^{+    16.1}_{-    18.5} $&$^{+     2.2}_{-     7.4}$
&   98.0 &$ ^{+     1.9}_{-     5.1} $&$^{+     4.7}_{-     2.5} $\\
\hline
$ 200 $&    39.6 &$ ^{+    15.9}_{-    18.4} $&$^{+     1.8}_{-     7.8}$
&   82.7 &$ ^{+     1.7}_{-     4.4} $&$^{+     5.2}_{-     2.4} $\\
\hline
$ 210 $&    32.5 &$ ^{+    15.9}_{-    18.0} $&$^{+     2.9}_{-     7.6}$
&   70.7 &$ ^{+     1.6}_{-     4.8} $&$^{+     4.9}_{-     2.5} $\\
\hline
$ 220 $&    26.5 &$ ^{+    16.0}_{-    17.6} $&$^{+     3.9}_{-     7.5}$
&   60.4 &$ ^{+     1.5}_{-     5.1} $&$^{+     4.6}_{-     2.6} $\\
\hline
$ 230 $&    21.7 &$ ^{+    16.0}_{-    17.2} $&$^{+     5.0}_{-     7.3}$
&   52.0 &$ ^{+     1.3}_{-     5.4} $&$^{+     4.3}_{-     2.7} $\\
\hline
$ 240 $&    18.1 &$ ^{+    16.1}_{-    16.8} $&$^{+     6.1}_{-     7.1}$
&   44.9 &$ ^{+     1.2}_{-     5.8} $&$^{+     4.0}_{-     2.8} $\\
\hline
$ 250 $&    15.1 &$ ^{+    16.1}_{-    16.4} $&$^{+     7.1}_{-     6.9}$
&   39.0 &$ ^{+     1.1}_{-     6.1} $&$^{+     3.6}_{-     2.9} $\\
\hline
$ 260 $&    12.7 &$ ^{+    16.0}_{-    16.8} $&$^{+     7.1}_{-     7.6}$
&   33.9 &$ ^{+     1.4}_{-     6.5} $&$^{+     3.6}_{-     3.3} $\\
\hline
$ 270 $&    10.8 &$ ^{+    15.8}_{-    17.1} $&$^{+     7.0}_{-     8.2}$
&   29.7 &$ ^{+     1.6}_{-     6.9} $&$^{+     3.6}_{-     3.8} $\\
\hline
$ 280 $&    9.09 &$ ^{+    15.7}_{-    17.4} $&$^{+     7.0}_{-     8.9}$
&   26.0 &$ ^{+     1.9}_{-     7.3} $&$^{+     3.6}_{-     4.2} $\\
\hline
$ 290 $&    7.83 &$ ^{+    15.5}_{-    17.7} $&$^{+     6.9}_{-     9.5}$
&   22.9 &$ ^{+     2.2}_{-     7.7} $&$^{+     3.6}_{-     4.6} $\\
\hline
$ 300 $&    6.68 &$ ^{+    15.4}_{-    18.0} $&$^{+     6.8}_{-    10.2}$
&   20.2 &$ ^{+     2.5}_{-     8.1} $&$^{+     3.5}_{-     5.1} $\\
\hline
\end{tabular}

      \caption[]{\label{tab::jety48-14}Central values, scale and
        \pdf{} uncertainties for the exclusive $H+0$-jet (\nnlo) and
        inclusive $H+$jet (\nlo) cross section for
        $|y^{\text{jet}}| < 2.5$, see also \fig{fig::jetmH-14}\,(a).}
\end{table}


\begin{table}[h!tbp]
  \centering
      \renewcommand{\arraystretch}{1.2}
\begin{tabular}{|c||c|c|c||c|c|c|}
\hline\multicolumn{7}{|c|}{\lhc{} @ 14\,TeV,\ \ $|y^\text{jet}|<2.5$,\ \ $p_T^\text{jet} > 20$\,GeV,\ \ $R = 0.4$}
\\\hline
$\mhiggs{}$&\multicolumn{1}{|c|}{$\sigma_{0\text{-jet}}^\text{\nnlo}$}&
scale&\pdf{}&
\multicolumn{1}{|c|}{$\sigma_{\geq1\text{-jet}}^{\text{\nlo}}$}&scale&\pdf{}\\[-.4em]

 \footnotesize[GeV]&\multicolumn{1}{|c|}{\footnotesize[fb]}&\footnotesize[\%]&\footnotesize[\%]&\multicolumn{1}{|c|}{\footnotesize[fb]}&\footnotesize[\%]&\footnotesize[\%]\\\hline\hline
$ 100 $& 673 &$ ^{+    10.2}_{-    18.8} $&$^{+     2.7}_{-     3.7}$
&514 &$ ^{+     8.4}_{-    13.5} $&$^{+     3.1}_{-     4.3} $\\
\hline
$ 110 $& 487 &$ ^{+    10.4}_{-    17.8} $&$^{+     2.9}_{-     3.3}$
&401 &$ ^{+     7.1}_{-    12.2} $&$^{+     3.3}_{-     4.0} $\\
\hline
$ 120 $& 360 &$ ^{+    10.5}_{-    16.8} $&$^{+     3.1}_{-     3.0}$
&317 &$ ^{+     5.7}_{-    10.8} $&$^{+     3.4}_{-     3.7} $\\
\hline
$ 130 $& 271 &$ ^{+    10.7}_{-    15.8} $&$^{+     3.3}_{-     2.6}$
&254 &$ ^{+     4.3}_{-     9.5} $&$^{+     3.6}_{-     3.4} $\\
\hline
$ 140 $& 207 &$ ^{+    10.9}_{-    14.8} $&$^{+     3.5}_{-     2.2}$
&206 &$ ^{+     3.0}_{-     8.2} $&$^{+     3.8}_{-     3.1} $\\
\hline
$ 150 $& 161 &$ ^{+    11.1}_{-    13.9} $&$^{+     3.7}_{-     1.8}$
&168 &$ ^{+     1.6}_{-     6.9} $&$^{+     3.9}_{-     2.7} $\\
\hline
$ 160 $& 126 &$ ^{+    11.0}_{-    13.4} $&$^{+     3.7}_{-     2.6}$
&139 &$ ^{+     1.5}_{-     6.6} $&$^{+     3.9}_{-     2.7} $\\
\hline
$ 170 $& 100 &$ ^{+    10.8}_{-    12.9} $&$^{+     3.6}_{-     3.4}$
&115 &$ ^{+     1.4}_{-     6.2} $&$^{+     3.9}_{-     2.7} $\\
\hline
$ 180 $&    80.5 &$ ^{+    10.7}_{-    12.4} $&$^{+     3.5}_{-     4.2}$
&   96.6 &$ ^{+     1.4}_{-     5.9} $&$^{+     3.9}_{-     2.7} $\\
\hline
$ 190 $&    65.0 &$ ^{+    10.6}_{-    12.0} $&$^{+     3.5}_{-     5.1}$
&   81.3 &$ ^{+     1.3}_{-     5.6} $&$^{+     3.8}_{-     2.7} $\\
\hline
$ 200 $&    53.2 &$ ^{+    10.4}_{-    11.5} $&$^{+     3.4}_{-     5.9}$
&   68.9 &$ ^{+     1.2}_{-     5.3} $&$^{+     3.8}_{-     2.6} $\\
\hline
$ 210 $&    43.7 &$ ^{+    10.5}_{-    11.3} $&$^{+     3.7}_{-     5.5}$
&   58.6 &$ ^{+     1.3}_{-     5.4} $&$^{+     3.7}_{-     3.1} $\\
\hline
$ 220 $&    36.2 &$ ^{+    10.5}_{-    11.1} $&$^{+     3.9}_{-     5.1}$
&   50.4 &$ ^{+     1.4}_{-     5.5} $&$^{+     3.5}_{-     3.6} $\\
\hline
$ 230 $&    30.1 &$ ^{+    10.5}_{-    10.9} $&$^{+     4.2}_{-     4.7}$
&   43.5 &$ ^{+     1.5}_{-     5.7} $&$^{+     3.4}_{-     4.0} $\\
\hline
$ 240 $&    25.3 &$ ^{+    10.6}_{-    10.8} $&$^{+     4.5}_{-     4.3}$
&   37.6 &$ ^{+     1.6}_{-     5.8} $&$^{+     3.2}_{-     4.5} $\\
\hline
$ 250 $&    21.4 &$ ^{+    10.6}_{-    10.6} $&$^{+     4.7}_{-     3.9}$
&   32.6 &$ ^{+     1.7}_{-     5.9} $&$^{+     3.1}_{-     5.0} $\\
\hline
$ 260 $&    18.1 &$ ^{+    10.3}_{-    10.4} $&$^{+     4.8}_{-     4.7}$
&   28.5 &$ ^{+     1.7}_{-     6.3} $&$^{+     3.2}_{-     5.0} $\\
\hline
$ 270 $&    15.5 &$ ^{+     9.9}_{-    10.1} $&$^{+     4.9}_{-     5.4}$
&   24.8 &$ ^{+     1.8}_{-     6.7} $&$^{+     3.4}_{-     5.1} $\\
\hline
$ 280 $&    13.2 &$ ^{+     9.5}_{-     9.9} $&$^{+     5.0}_{-     6.2}$
&   21.8 &$ ^{+     1.9}_{-     7.0} $&$^{+     3.5}_{-     5.2} $\\
\hline
$ 290 $&    11.4 &$ ^{+     9.2}_{-     9.7} $&$^{+     5.0}_{-     6.9}$
&   19.3 &$ ^{+     2.0}_{-     7.4} $&$^{+     3.7}_{-     5.2} $\\
\hline
$ 300 $&    9.73 &$ ^{+     8.8}_{-     9.5} $&$^{+     5.1}_{-     7.7}$
&   17.0 &$ ^{+     2.1}_{-     7.8} $&$^{+     3.8}_{-     5.3} $\\
\hline
\end{tabular}

	  \caption{\label{tab::jety25-14}Same as
            Table\,\ref{tab::jety48-14}, but for $|y^{\text{jet}}| < 2.5$,
            see also \fig{fig::jetmH-14}\,(b).}
\end{table}


\begin{table}[h!tbp]
  \centering
      \renewcommand{\arraystretch}{1.2}
\begin{tabular}{|c||c|c||c|c||c|c|}
\hline\multicolumn{7}{|c|}{\lhc{} @ 14\,TeV}
\\\hline
&&&
\multicolumn{4}{|c|}{$p_T^\text{jet}>20$\,GeV,\ \ $R = 0.4$}\\
\cline{4-7}
&&&
\multicolumn{2}{|c||}{$|y^\text{jet}|<4.8$}
&
\multicolumn{2}{|c|}{$|y^\text{jet}|<2.5$}
\\\cline{4-7}
$\mhiggs{}$
&
\multicolumn{1}{|c|}{$\sigma_\text{tot}^\text{\nnlo}$}
&
scale
&
\multicolumn{1}{|c|}{$\sigma_{\geq1\text{-jet}}^{\text{\nlo}'}$}
&
scale
&
\multicolumn{1}{|c|}{$\sigma_{\geq1\text{-jet}}^{\text{\nlo}'}$}
&
scale
\\[-.4em]

 \footnotesize[GeV]
&
\multicolumn{1}{|c|}{\footnotesize[fb]}
&
\footnotesize[\%]
&
\multicolumn{1}{|c|}{\footnotesize[fb]}
&
\footnotesize[\%]
&
\multicolumn{1}{|c|}{\footnotesize[fb]}
&
\footnotesize[\%]
\\\hline\hline
$ 100 $& 1178 &$^{+     1.2}_{-     9.5}$&
620&$^{+     8.2}_{-    12.7}$&505&$^{+     8.6}_{-    12.1}$\\\hline
$ 110 $& 882 &$^{+     1.3}_{-     8.8}$&
482&$^{+     6.9}_{-    11.7}$&395&$^{+     7.1}_{-    11.1}$\\\hline
$ 120 $& 673 &$^{+     1.4}_{-     8.1}$&
381&$^{+     5.6}_{-    10.6}$&313&$^{+     5.5}_{-    10.1}$\\\hline
$ 130 $& 522 &$^{+     1.5}_{-     7.4}$&
305&$^{+     4.3}_{-     9.6}$&251&$^{+     4.0}_{-     9.1}$\\\hline
$ 140 $& 411 &$^{+     1.7}_{-     6.7}$&
246&$^{+     3.0}_{-     8.5}$&203&$^{+     2.4}_{-     8.1}$\\\hline
$ 150 $& 327 &$^{+     1.8}_{-     6.0}$&
201&$^{+     1.7}_{-     7.5}$&166&$^{+     0.9}_{-     7.1}$\\\hline
$ 160 $& 263 &$^{+     1.8}_{-     5.9}$&
166&$^{+     1.8}_{-     6.9}$&137&$^{+     0.9}_{-     6.6}$\\\hline
$ 170 $& 214 &$^{+     1.9}_{-     5.8}$&
137&$^{+     1.9}_{-     6.3}$&114&$^{+     0.9}_{-     6.1}$\\\hline
$ 180 $& 176 &$^{+     2.0}_{-     5.8}$&
115&$^{+     2.0}_{-     5.8}$&   95.5&$^{+     0.9}_{-     5.5}$\\\hline
$ 190 $& 146 &$^{+     2.0}_{-     5.7}$&
   96.7&$^{+     2.1}_{-     5.2}$&   80.7&$^{+     0.9}_{-     5.0}$\\\hline
$ 200 $& 121 &$^{+     2.1}_{-     5.6}$&
   81.9&$^{+     2.2}_{-     4.6}$&   68.3&$^{+     0.9}_{-     4.5}$\\\hline
$ 210 $& 102 &$^{+     2.1}_{-     5.5}$&
   69.5&$^{+     2.1}_{-     4.8}$&   58.3&$^{+     1.0}_{-     4.7}$\\\hline
$ 220 $&    86.2 &$^{+     2.1}_{-     5.5}$&
   59.7&$^{+     1.9}_{-     5.1}$&   50.1&$^{+     1.0}_{-     5.0}$\\\hline
$ 230 $&    73.3 &$^{+     2.1}_{-     5.4}$&
   51.6&$^{+     1.8}_{-     5.3}$&   43.2&$^{+     1.0}_{-     5.2}$\\\hline
$ 240 $&    62.7 &$^{+     2.1}_{-     5.3}$&
   44.6&$^{+     1.7}_{-     5.5}$&   37.4&$^{+     1.1}_{-     5.4}$\\\hline
$ 250 $&    53.8 &$^{+     2.1}_{-     5.3}$&
   38.7&$^{+     1.5}_{-     5.7}$&   32.4&$^{+     1.1}_{-     5.7}$\\\hline
$ 260 $&    46.4 &$^{+     2.1}_{-     5.2}$&
   33.7&$^{+     1.6}_{-     6.1}$&   28.3&$^{+     1.1}_{-     5.9}$\\\hline
$ 270 $&    40.2 &$^{+     2.1}_{-     5.1}$&
   29.5&$^{+     1.8}_{-     6.5}$&   24.7&$^{+     1.0}_{-     6.1}$\\\hline
$ 280 $&    35.0 &$^{+     2.1}_{-     5.1}$&
   25.9&$^{+     1.9}_{-     6.8}$&   21.8&$^{+     1.0}_{-     6.2}$\\\hline
$ 290 $&    30.6 &$^{+     2.1}_{-     5.0}$&
   22.7&$^{+     2.0}_{-     7.2}$&   19.2&$^{+     1.0}_{-     6.4}$\\\hline
$ 300 $&    26.8 &$^{+     2.1}_{-     4.9}$&
   20.1&$^{+     2.1}_{-     7.6}$&   17.1&$^{+     0.9}_{-     6.6}$\\\hline
\end{tabular}

	  \caption[]{\label{tab::tot12p-14}Total cross section
            $\sigma_\text{tot}^\nnlo$, and
            $\sigma_{\geq1\text{-jet}}^{\nlo'}$ for two different jet
            rapidity cuts, including the associated scale uncertainties.}
\end{table}


\begin{figure}[h!tb]
  \begin{center}
    \begin{tabular}{c}
      \subfigure[]{%
        \includegraphics[width=.45\textwidth]{%
          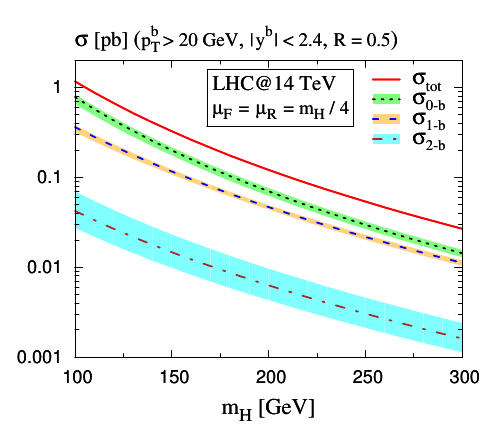}}
      \subfigure[]{%
        \includegraphics[width=.45\textwidth]{%
          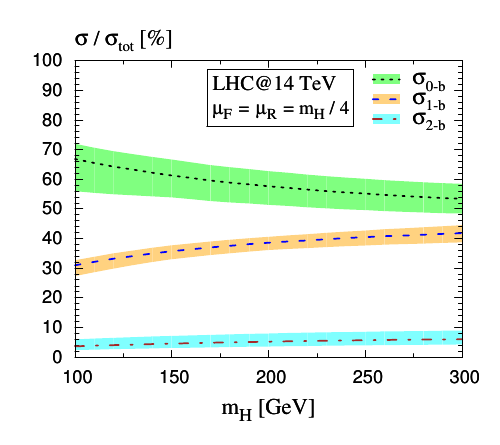}}
    \end{tabular}
    \caption[]{Higgs mass dependence of the $H+nb$-jet contributions
      with respect to the total cross section (a)~in absolut numbers and
      (b)~relative to the total cross section.}
    \label{fig::bjetmH-14}
  \end{center}
\end{figure}


\begin{table}[h!tbp]
  \centering
      \renewcommand{\arraystretch}{1.2}
\begin{tabular}{|c||c|c|c||c|c|c||c|c|c|}
\hline\multicolumn{10}{|c|}{\lhc{} @ 14\,TeV,\ \ $|y^{b}|<2.4$,\ \ $p_T^b > 20$\,GeV,\ \ $R = 0.5$}
\\\hline
$\mhiggs{}$&\multicolumn{1}{|c|}{$\sigma_{0b}^\text{\nnlo{}}$}&
scale&
\pdf{}&
\multicolumn{1}{|c|}{$\sigma_{1b}^\text{\nlo}$}&
scale&
\pdf{}&
\multicolumn{1}{|c|}{$\sigma_{2b}^\text{\lo}$}&
scale&
\pdf{}\\[-.4em]
\footnotesize[GeV]&\multicolumn{1}{|c|}{\footnotesize[fb]}&\footnotesize[\%]&\footnotesize[\%]&\multicolumn{1}{|c|}{\footnotesize[fb]}&\footnotesize[\%]&\footnotesize[\%]&\multicolumn{1}{|c|}{\footnotesize[fb]}&\footnotesize[\%]&\footnotesize[\%]\\\hline\hline
100&787&$^{+     7.1}_{-    16.0} $&$^{+     3.0}_{-     2.4}$&
364 &$ ^{+     4.4}_{-    11.4} $&$^{+     3.2}_{-     2.7}$&
   42.3 &$ ^{+    64.7}_{-    34.8} $&$^{+     1.8}_{-     2.5}$
 \\
\hline
110&577&$^{+     7.3}_{-    15.2} $&$^{+     3.0}_{-     2.6}$&
283 &$ ^{+     4.3}_{-    10.4} $&$^{+     3.4}_{-     2.9}$&
   33.4 &$ ^{+    63.3}_{-    34.4} $&$^{+     1.8}_{-     2.5}$
 \\
\hline
120&432&$^{+     7.6}_{-    14.3} $&$^{+     2.9}_{-     2.8}$&
223 &$ ^{+     4.1}_{-     9.5} $&$^{+     3.6}_{-     3.1}$&
   26.8 &$ ^{+    61.8}_{-    34.0} $&$^{+     1.7}_{-     2.6}$
 \\
\hline
130&330&$^{+     7.8}_{-    13.4} $&$^{+     2.9}_{-     2.9}$&
178 &$ ^{+     4.0}_{-     8.5} $&$^{+     3.8}_{-     3.2}$&
   21.7 &$ ^{+    60.4}_{-    33.6} $&$^{+     1.7}_{-     2.6}$
 \\
\hline
140&255&$^{+     8.1}_{-    12.5} $&$^{+     2.8}_{-     3.1}$&
143 &$ ^{+     3.9}_{-     7.6} $&$^{+     3.9}_{-     3.4}$&
   17.8 &$ ^{+    59.0}_{-    33.1} $&$^{+     1.7}_{-     2.7}$
 \\
\hline
150&200&$^{+     8.3}_{-    11.6} $&$^{+     2.7}_{-     3.3}$&
117 &$ ^{+     3.7}_{-     6.6} $&$^{+     4.1}_{-     3.5}$&
   14.7 &$ ^{+    57.6}_{-    32.7} $&$^{+     1.7}_{-     2.7}$
 \\
\hline
160&159&$^{+     8.3}_{-    11.4} $&$^{+     2.8}_{-     3.2}$&
   95.7 &$ ^{+     3.7}_{-     6.3} $&$^{+     4.0}_{-     3.6}$&
   12.2 &$ ^{+    56.8}_{-    32.4} $&$^{+     1.8}_{-     2.7}$
 \\
\hline
170&128&$^{+     8.3}_{-    11.2} $&$^{+     2.9}_{-     3.0}$&
   79.2 &$ ^{+     3.7}_{-     5.9} $&$^{+     3.9}_{-     3.7}$&
   10.3 &$ ^{+    55.9}_{-    32.1} $&$^{+     1.9}_{-     2.6}$
 \\
\hline
180&104&$^{+     8.3}_{-    11.0} $&$^{+     3.0}_{-     2.9}$&
   66.0 &$ ^{+     3.8}_{-     5.6} $&$^{+     3.8}_{-     3.8}$&
   8.66 &$ ^{+    55.1}_{-    31.9} $&$^{+     2.0}_{-     2.6}$
 \\
\hline
190&   84.9&$^{+     8.3}_{-    10.8} $&$^{+     3.1}_{-     2.7}$&
   55.5 &$ ^{+     3.8}_{-     5.2} $&$^{+     3.7}_{-     3.9}$&
   7.35 &$ ^{+    54.2}_{-    31.6} $&$^{+     2.0}_{-     2.5}$
 \\
\hline
200&   70.0&$^{+     8.3}_{-    10.6} $&$^{+     3.1}_{-     2.6}$&
   46.8 &$ ^{+     3.8}_{-     4.9} $&$^{+     3.6}_{-     4.0}$&
   6.27 &$ ^{+    53.4}_{-    31.3} $&$^{+     2.1}_{-     2.5}$
 \\
\hline
210&   58.3&$^{+     8.3}_{-    10.3} $&$^{+     3.2}_{-     2.8}$&
   39.8 &$ ^{+     3.8}_{-     5.1} $&$^{+     3.7}_{-     4.0}$&
   5.38 &$ ^{+    52.8}_{-    31.1} $&$^{+     2.1}_{-     2.5}$
 \\
\hline
220&   48.7&$^{+     8.3}_{-    10.1} $&$^{+     3.2}_{-     3.0}$&
   33.9 &$ ^{+     3.9}_{-     5.2} $&$^{+     3.7}_{-     3.9}$&
   4.64 &$ ^{+    52.3}_{-    30.9} $&$^{+     2.1}_{-     2.5}$
 \\
\hline
230&   41.0&$^{+     8.4}_{-     9.9} $&$^{+     3.3}_{-     3.3}$&
   29.1 &$ ^{+     3.9}_{-     5.4} $&$^{+     3.8}_{-     3.9}$&
   4.01 &$ ^{+    51.7}_{-    30.7} $&$^{+     2.1}_{-     2.6}$
 \\
\hline
240&   34.8&$^{+     8.4}_{-     9.7} $&$^{+     3.3}_{-     3.5}$&
   25.1 &$ ^{+     4.0}_{-     5.6} $&$^{+     3.8}_{-     3.9}$&
   3.48 &$ ^{+    51.2}_{-    30.5} $&$^{+     2.1}_{-     2.6}$
 \\
\hline
250&   29.7&$^{+     8.4}_{-     9.4} $&$^{+     3.3}_{-     3.7}$&
   21.8 &$ ^{+     4.0}_{-     5.7} $&$^{+     3.9}_{-     3.8}$&
   3.04 &$ ^{+    50.6}_{-    30.4} $&$^{+     2.1}_{-     2.6}$
 \\
\hline
260&   25.4&$^{+     8.4}_{-     9.2} $&$^{+     3.4}_{-     3.8}$&
   18.9 &$ ^{+     4.1}_{-     5.9} $&$^{+     4.0}_{-     3.8}$&
   2.66 &$ ^{+    50.2}_{-    30.2} $&$^{+     2.2}_{-     2.7}$
 \\
\hline
270&   21.9&$^{+     8.5}_{-     9.0} $&$^{+     3.6}_{-     3.9}$&
   16.5 &$ ^{+     4.1}_{-     6.1} $&$^{+     4.2}_{-     3.7}$&
   2.33 &$ ^{+    49.8}_{-    30.0} $&$^{+     2.2}_{-     2.7}$
 \\
\hline
280&   18.9&$^{+     8.5}_{-     8.8} $&$^{+     3.7}_{-     4.0}$&
   14.4 &$ ^{+     4.2}_{-     6.2} $&$^{+     4.4}_{-     3.7}$&
   2.05 &$ ^{+    49.3}_{-    29.9} $&$^{+     2.3}_{-     2.7}$
 \\
\hline
290&   16.4&$^{+     8.5}_{-     8.6} $&$^{+     3.8}_{-     4.1}$&
   12.7 &$ ^{+     4.3}_{-     6.4} $&$^{+     4.6}_{-     3.6}$&
   1.81 &$ ^{+    48.9}_{-    29.7} $&$^{+     2.3}_{-     2.7}$
 \\
\hline
300&   14.3&$^{+     8.6}_{-     8.4} $&$^{+     3.9}_{-     4.2}$&
   11.2 &$ ^{+     4.4}_{-     6.6} $&$^{+     4.7}_{-     3.5}$&
   1.60 &$ ^{+    48.5}_{-    29.6} $&$^{+     2.4}_{-     2.8}$
 \\
\hline
\end{tabular}

	  \caption{\label{tab::012blhc14}Central values, scale and \pdf{}
            uncertainties for the exclusive $H+nb$-jet ($n=0,1,2$) cross
            section at \nnlo{}, \nlo{}, and \lo{}, respectively. See
            also \fig{fig::bjetmH-14}.}
\end{table}



\end{document}